\documentclass[aps,prc,superscriptaddress,floatfix,nofootinbib]{revtex4}
\usepackage{color}
\usepackage{dcolumn}
\usepackage{graphicx}

\begin{document}

\title{Energy dependence of $\phi$ meson production \\
in central Pb+Pb collisions 
at $\sqrt{s_{NN}}$ = 6 to 17 GeV}


\affiliation{NIKHEF, Amsterdam, Netherlands}  
\affiliation{Department of Physics, University of Athens, Athens, Greece}
\affiliation{Comenius University, Bratislava, Slovakia}
\affiliation{KFKI Research Institute for Particle and Nuclear Physics,
             Budapest, Hungary}
\affiliation{MIT, Cambridge, Massachusetts, USA}
\affiliation{Henryk Niewodniczanski Institute of Nuclear Physics,
             Polish Academy of Science, Cracow, Poland}
\affiliation{Gesellschaft f\"{u}r Schwerionenforschung (GSI),
             Darmstadt, Germany} 
\affiliation{Joint Institute for Nuclear Research, Dubna, Russia}
\affiliation{Fachbereich Physik der Universit\"{a}t, Frankfurt, Germany}
\affiliation{CERN, Geneva, Switzerland}
\affiliation{Institute of Physics \'Swi{\,e}tokrzyska Academy, Kielce, Poland}
\affiliation{Fachbereich Physik der Universit\"{a}t, Marburg, Germany}
\affiliation{Max-Planck-Institut f\"{u}r Physik, Munich, Germany}
\affiliation{Institute of Particle and Nuclear Physics, Charles
             University, Prague, Czech Republic}
\affiliation{Department of Physics, Pusan National University, Pusan,
             Republic of Korea} 
\affiliation{Nuclear Physics Laboratory, University of Washington,
             Seattle, Washington, USA} 
\affiliation{Atomic Physics Department, Sofia University St.~Kliment
             Ohridski, Sofia, Bulgaria} 
\affiliation{Institute for Nuclear Research and Nuclear Energy, Sofia, Bulgaria}
\affiliation{Department of Chemistry, Stony Brook University (SUNYSB), Stony Brook, New York, USA}
\affiliation{Institute for Nuclear Studies, Warsaw, Poland}
\affiliation{Institute for Experimental Physics, University of Warsaw,
             Warsaw, Poland} 
\affiliation{Faculty of Physics, Warsaw University of Technology, Warsaw, Poland}
\affiliation{Rudjer Boskovic Institute, Zagreb, Croatia}


\author{C.~Alt}
\affiliation{Fachbereich Physik der Universit\"{a}t, Frankfurt, Germany}
\author{T.~Anticic} 
\affiliation{Rudjer Boskovic Institute, Zagreb, Croatia}
\author{B.~Baatar}
\affiliation{Joint Institute for Nuclear Research, Dubna, Russia}
\author{D.~Barna}
\affiliation{KFKI Research Institute for Particle and Nuclear Physics,
             Budapest, Hungary} 
\author{J.~Bartke}
\affiliation{Henryk Niewodniczanski Institute of Nuclear Physics,
             Polish Academy of Science, Cracow, Poland}
\author{L.~Betev}
\affiliation{CERN, Geneva, Switzerland}
\author{H.~Bia{\l}\-kowska} 
\affiliation{Institute for Nuclear Studies, Warsaw, Poland}
\author{C.~Blume}
\affiliation{Fachbereich Physik der Universit\"{a}t, Frankfurt, Germany}
\author{B.~Boimska}
\affiliation{Institute for Nuclear Studies, Warsaw, Poland}
\author{M.~Botje}
\affiliation{NIKHEF, Amsterdam, Netherlands}
% ...


\collaboration{NA49 Collaboration}
\noaffiliation


\begin{abstract}

$\phi$ meson production is studied by the NA49 Collaboration in central
Pb+Pb collisions at 20$A$, 30$A$, 40$A$,
80$A$, and 158$A$~GeV beam energy. The data are compared with measurements
at lower and higher energies and with microscopic and thermal
models. The energy dependence of yields and spectral distributions
is compatible with the assumption that partonic degrees of freedom 
set in at low SPS energies.

\end{abstract}

\maketitle

\section{Introduction}

The production of strange particles is considered one of the 
key observables for 
understanding the reaction mechanisms in ultrarelativistic heavy-ion
collisions. Enhanced strangeness production with respect to
proton-proton collisions was originally proposed as a signature of 
the transition to a deconfined state of quarks and gluons
during the initial stage of the reactions~\cite{koch1986}.
The enhancement was predicted to arise from gluon fragmentation into 
quark-antiquark pairs which is believed to have a significantly lower
threshold than strange-antistrange hadron pair production channels.
Indeed, it has been observed~\cite{na35,sikler1999} that the ratio of the number 
of produced kaons to that of pions is higher by a factor of about 2 in central S + S and Pb + Pb
reactions than that in $p+p$ collisions at the top energy available at the
CERN Super Proton Synchroton (SPS). 

Statistical hadron gas models have been successfully employed to describe
the measured particle yields at various collision 
energies~\cite{cleymans1999,pbm1999,pbm2001,averbeck2003,becattini2006}. 
The fact that the hadronic final state of the collision resembles a hadron gas
in chemical equilibrium has been interpreted as a consequence of the
hadronization process~\cite{stock1999} or as a result of a fast hadronic equilibration
process involving multiparticle collisions~\cite{pbm2004}. In this
hadron gas picture, enhanced production of strange particles in collisions
of large nuclei arises as a consequence of the increased reaction volume, 
relaxing the influence of strangeness conservation~\cite{hagedorn1985}.
Technically, this requires the application of the canonical ensemble to
small collision systems, while for larger volumes such as those encountered in
central collisions of heavy ions, the grand-canonical approximation
is valid. It has been shown that this ``canonical strangeness suppression'' 
also applies to a partonic system~\cite{rafelski1980}.

In addition to this volume effect, the strange particle phase space
appears to be undersaturated in elementary interactions. The deviation
of the strange particle yields from a hadron gas in full equilibrium
was parametrized by a strangeness undersaturation factor 
$\gamma_S$~\cite{becattini1997,becattini2006}. 
The additional suppression becomes much weaker in heavy-ion collisions.
However, fits to the hadron multiplicities in full phase space 
are still unsatisfactory when not taking into account $\gamma_S$
\cite{becattini2006}. A possible
interpretation is that the total amount of strangeness available
for hadronization is determined in a prehadronic stage of the
collision. A change in $\gamma_S$ between $p+p$ and $A+A$ would then
reflect the difference in the initial conditions of the respective
fireballs.

The hadron gas model was extended to describe the energy dependence
of produced hadron multiplicities by a smooth parametrization of the fit
parameters $T$ and $\mu_B$, determined at energies available at the BNL
Alternating Gradient Synchroton (AGS), SPS, and BNL Relativistic Heavy Ion Collider (RHIC),
as a function of collision energy~\cite{pbm2002}.
However, this extended model failed
to reproduce 
the detailed features of the energy dependence of relative strangeness
production measured
by NA49 in its energy scan program. In particular,
the sharp maximum at around 30$A$~GeV beam energy~\cite{gazdzicki2004, friese2005}
could not be described. 
The same holds true for microscopic reaction models such as UrQMD~\cite{urqmd}.
On the other hand, this feature was predicted as
a consequence of the onset of a phase transition to a deconfined state
at the respective beam energy~\cite{gazdzicki1999}. 

In this context, it is certainly interesting to
investigate specific strangeness-carrying hadrons.
Among these, the $\phi$ meson is of particular interest because of
its $\rm s \overline{s}$ valence quark composition. In a purely
hadronic scenario, being strangeness-neutral, it should not 
be sensitive to hadrochemical effects related to strangeness.
If on the other hand, the amount of available strange quarks is 
determined in a partonic stage of the collision, the $\phi$ is 
expected to react more sensitively than singly strange particles.
In particular, one would expect the $\phi$ meson yield to be suppressed
by $\gamma_s^2$ with respect to equilibrium. Analogously, the
canonical suppression mechanism in small systems should have a stronger 
effect on the $\phi$, leading to a larger relative enhancement in
Pb + Pb collisions with respect to $p+p$ reactions than observed for kaons. 

In the evolution of the fireball after hadronization, $\phi$ mesons can be
both formed by kaon coalescence and destroyed by rescattering.
In addition, when decaying inside the fireball, the daughter particles
can rescatter, leading to a loss of signal in the invariant mass peak
of the respective decay channel. This is more likely to happen
for slow $\phi$ mesons, which spend more time in the
fireball. Thus the effect could lead to a depletion of the
$\phi$ meson yield at low $p_t$ in central nucleus-nucleus collisions \cite{johnson2001}.

Theoretical investigations have suggested that the 
properties of the $\phi$ meson might be modified in a dense hadronic medium.
In particular, a decrease of its mass of the order of 
10~MeV\footnote{For better readability, we use natural units, i.e., $c=1$, throughout this article.}~\cite{hatsuda1992}
and an increase of its width by a factor of 2--3~\cite{lissauer1991} were predicted.
So far, there is only one experimental claim for a broadening of the width
in $p$ + Cu collisions~\cite{muto2007}.

In an earlier publication~\cite{na492000}, we reported on $\phi$ 
production at top SPS energy, where we found the $\phi$ enhanced by a factor of
about 3, compared to minimum bias $p+p$ collisions at the same
beam energy. Meanwhile, the $\phi$ meson was measured at the same energy
by the NA50~\cite{alessandro2003}, NA45~\cite{adamova2006}, and NA60~\cite{falco2006}
experiments. 
At the AGS, data on $\phi$ production were obtained by the E917 Collaboration in
Au + Au collisions at $p_{\rm beam} = 11.7 A$ GeV ($\sqrt{s_{NN}} = 4.88$ GeV)
in a restricted rapidity range $[y_{\rm c.m.}-0.4,y_{\rm c.m.}]$~\cite{back2004}.
At the RHIC, the STAR Collaboration measured the $\phi$ meson at 
$\sqrt{s_{NN}} = 130$ and $\sqrt{s_{NN}} = 200$~GeV
at midrapidity~\cite{adler2003,adams2005}. 
For the latter energy, data are also available from the PHENIX 
experiment~\cite{adler2005}.

In this article, we report on $\phi$ production in central Pb+Pb
collisions at five different beam energies from 20$A$ to 158$A$~GeV. Together
with the data obtained at the AGS and the RHIC,
our findings enable the study of
energy dependence of $\phi$ production over a large range
of collision energies.

\section{Experiment}
The NA49 experiment at CERN is based on a fixed-target hadron spectrometer
using heavy-ion beams from the SPS accelerator. Its main components
are four large-volume time projection chambers for charged-particle
tracking, two of which operate inside the magnetic field of two 
superconducting magnets, thus providing an excellent momentum measurement. 
Two larger main time projection chambers (MTPCs) are placed downstream, outside of the field, and
enable particle identification by the measurement of the specific
energy loss in the detector gas. The particle identification capabilities are enhanced
by a time-of-flight (TOF) scintillator system behind the MTPCs, albeit in
a restricted geometrical acceptance.

A thin lead foil with $1 \%$ interaction probability
for Pb nuclei was used as a target. For the different runs, the magnetic field was
scaled proportionally to the beam energies in order to have similar
acceptance in the c.m.~system.

The centrality of the reactions was determined from the energy 
deposited by the beam spectators in the zero-degree calorimeter, 
placed 20 m downstream of the target. By setting an upper limit
on this energy, the online central trigger selected the 7.2\% most central 
collisions at $20A$--$80A$~GeV and the 10\% most central collisions at 158$A$~GeV. 
The latter data set was restricted to 5\% centrality 
in the offline analysis. The
corresponding mean numbers of wounded nucleons were obtained by Glauber
model calculations (see Table~\ref{tab:datasets}).
Details of the experimental apparatus can be found in Ref.~\cite{na49nim}.


\section{Data Analysis}

\subsection{Event and track selection}
Offline quality criteria were applied to the events selected by the online 
centrality trigger to suppress nontarget interactions, pileup, and
incorrectly reconstructed events. The cut variables include the
position and $\chi^2$ of the reconstructed vertex and the track
multiplicity. For the central data sets used in this analysis, however,
the impact of these quality cuts is marginal; only about 1\% of all 
events were rejected. Table~\ref{tab:datasets} shows the event
statistics used in the analysis for the five data sets.

The analysis was restricted to tracks reconstructed in the MTPCs
which could be assigned to the primary vertex. A minimal track length
of 2~m out of the maximal 4~m in the MTPCs was required to 
suppress ghost or split tracks and to 
ensure a good resolution in $dE/dx$. Detailed studies including 
reconstruction of simulated tracks embedded into real raw data events
showed that for such a selection of tracks, losses due to track 
reconstruction and high track density are negligible.

\begin{table*}
\caption{\label{tab:datasets}Characteristics of the data sets employed in the analysis.
The mean numbers of wounded nucleons $\langle N_w \rangle$ were obtained by Glauber
model calculations.}
\begin{center}
\begin{ruledtabular}
\begin{tabular}{cccccccc}
$E_{\rm beam}$ &
$\sqrt{s_{NN}}$  & 
$y_{\rm beam}$ &
Year &
Centrality &
$\langle N_w \rangle$ &
$N_{\rm events}$ & 
Momentum range  \\
 ($A$~GeV) & (GeV) & & & & & & (GeV) \\
\hline
 20 &   6.3 &  1.88  &  2002  & 7.2\% & $349 \pm 1 \pm 5$ 
& 352 309  &  2.0--23.0 \\
 30 &   7.6 &  2.08  &  2002  & 7.2\% & $349 \pm 1 \pm 5$ 
& 368 662  &  2.0--27.0 \\
 40 &   8.8 &  2.22  &  1999  & 7.2\% & $349 \pm 1 \pm 5$ 
& 586 768  &  2.0--27.0 \\
 80 &  12.3 &  2.57  &  2000  & 7.2\% & $349 \pm 1 \pm 5$ 
& 300 992  &  2.0--32.0 \\
158 &  17.3 &  2.91  &  1996  & 5.0\% & $362 \pm 1 \pm 5$ 
& 345 543  &  3.5--35.0 \\   
\end{tabular}
\end{ruledtabular}
\end{center}
\end{table*} 

\subsection{Selection of kaon candidates}

\begin{figure*}
\includegraphics[width=0.45\linewidth]{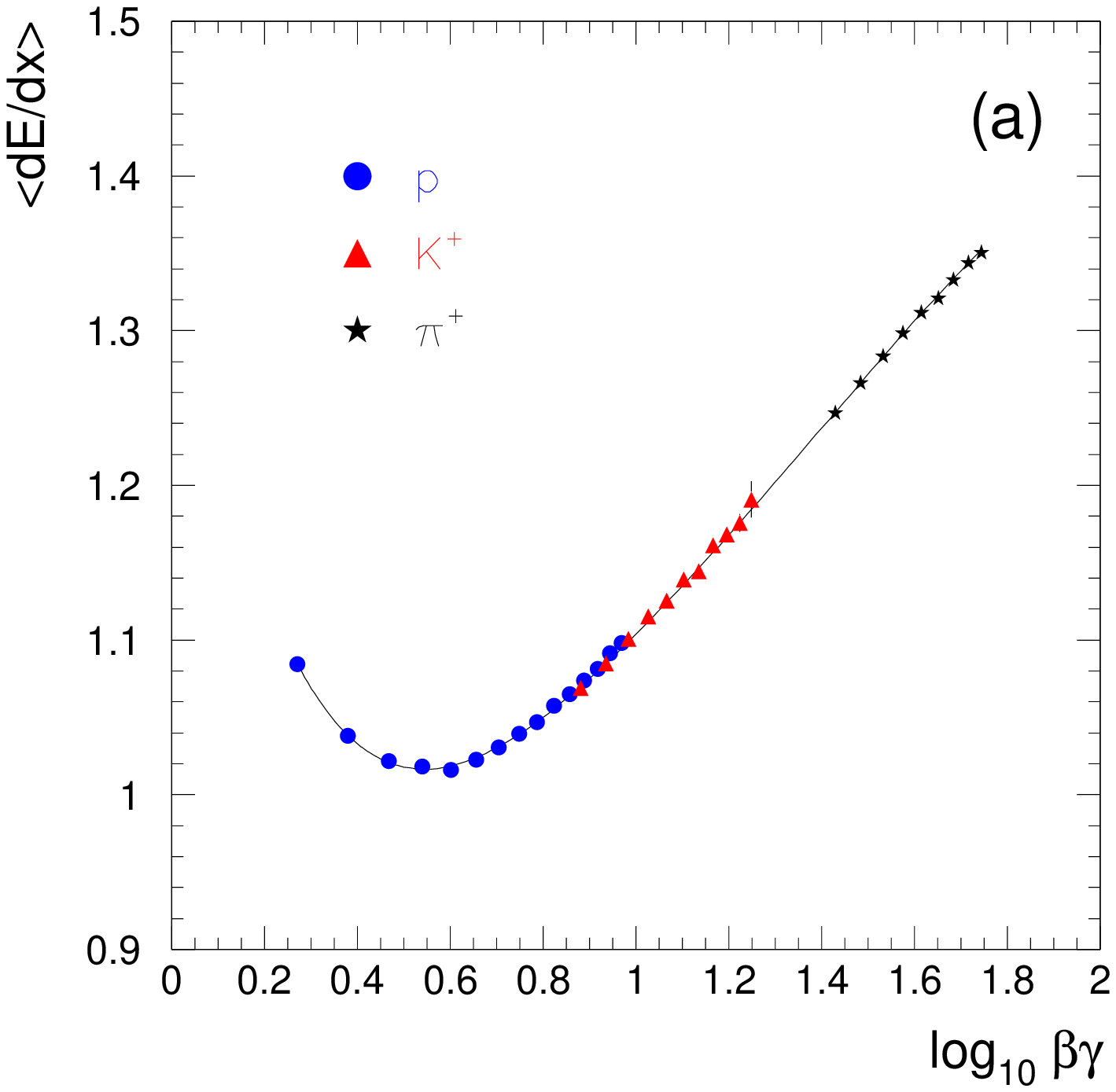}
\includegraphics[width=0.45\linewidth]{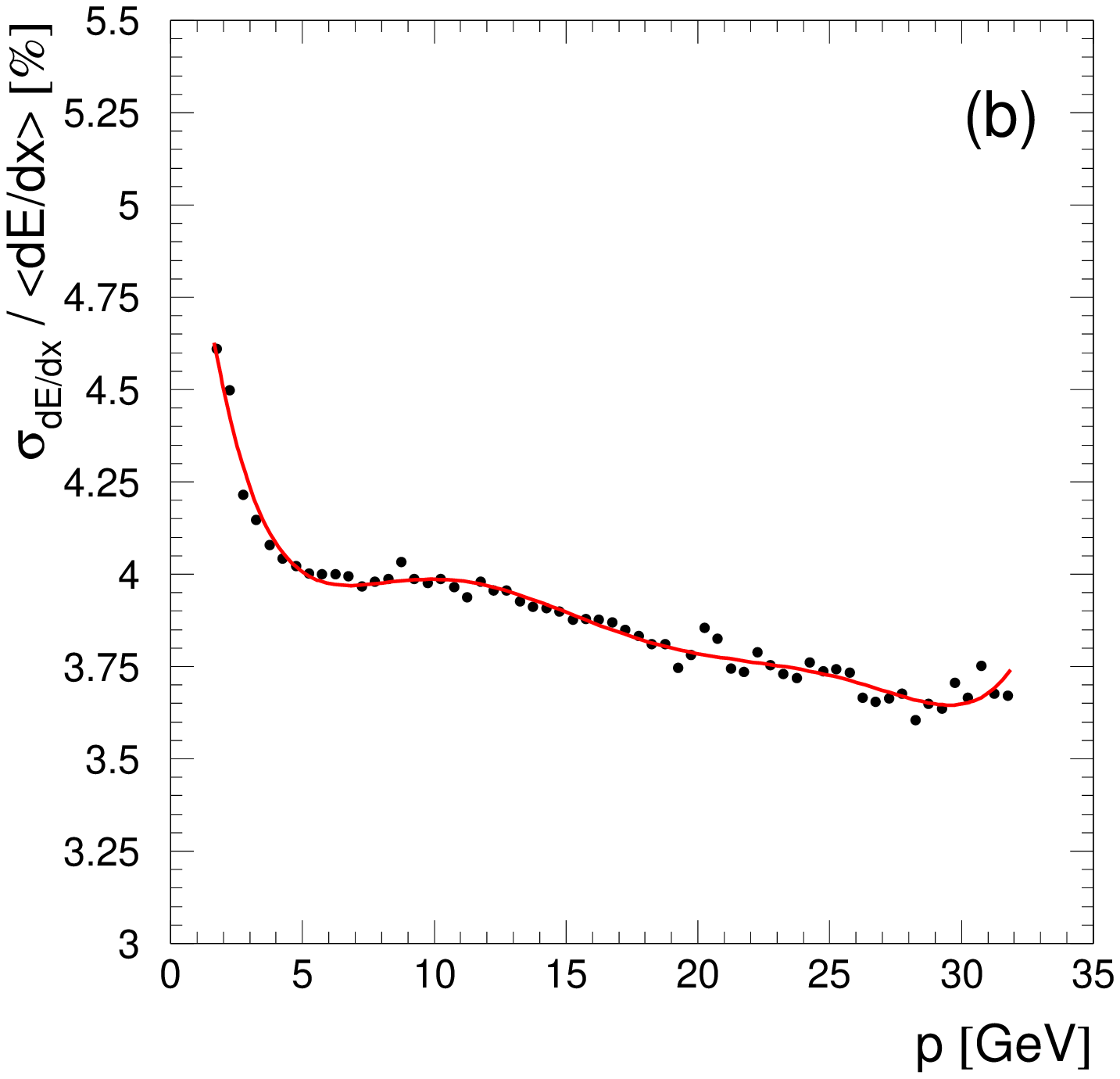}
\caption{(Color online) $dE/dx$ parametrization for the data set at 80$A$~GeV. (a)
Mean $dE/dx$ as function of $\beta\gamma$ determined for TOF-identified 
pions, kaons, and protons;
(b) $dE/dx$ resolution as function of momentum,
obtained from the deconvolution of the energy loss spectra into the 
contributions of $\pi^+$, $K^+$, and $p$.}
\label{fig:dedxparam}
\end{figure*}

NA49 observes the $\phi$ meson through its hadronic decay into charged kaons.
To reduce the large contribution of pions and protons to the 
combinatorial background, kaon candidates were selected based on their 
specific energy loss $dE/dx$ in the MTPCs. The mean $dE/dx$ of pions,
kaons, and (anti-)protons was determined from TOF-identified particles
in the acceptance of the time-of-flight detectors and parametrized 
as a function of $\beta\gamma$ as shown in Fig.~\ref{fig:dedxparam}(a).
This allowed one
to extend the momentum range for the identification from the TOF 
acceptance to higher momenta. The lower momentum limit was given 
by either the MTPC acceptance or the crossing of the Bethe-Bloch curves 
of pions and kaons. The momentum limits for the different data sets are 
summarized in Table~\ref{tab:datasets}.

Fixing the mean $dE/dx$ of kaons and protons to this
parametrization, the resolution was obtained by unfolding the
energy-loss spectra in momentum bins into the Gaussian contributions
of the particle species ($p$, $K$, $\pi$, and $e$).
The resolution is about 4\% and has a slight momentum dependence 
which was again parametrized [Fig.~\ref{fig:dedxparam}(b)].

Kaon candidates were selected by a momentum-dependent $dE/dx$ window 
around the expectation value, the size of which was chosen to optimize
the $\phi$ signal quality. In addition, the window had to
be symmetric and large enough to minimize the sensitivity to the errors 
in the determination of the $dE/dx$ expectation value and resolution. 
A window of $\pm 1.5 \, \sigma$ was found to be the best choice. This selection 
contains 87\% of all kaons, giving an efficiency of 
75\% for the pair. The fraction of true kaons within the selected
candidate track sample varies between 40\% and 60\%.

\subsection{Extraction of raw yields}

\begin{figure*}
\begin{center}
\leavevmode
\includegraphics[width=\linewidth]{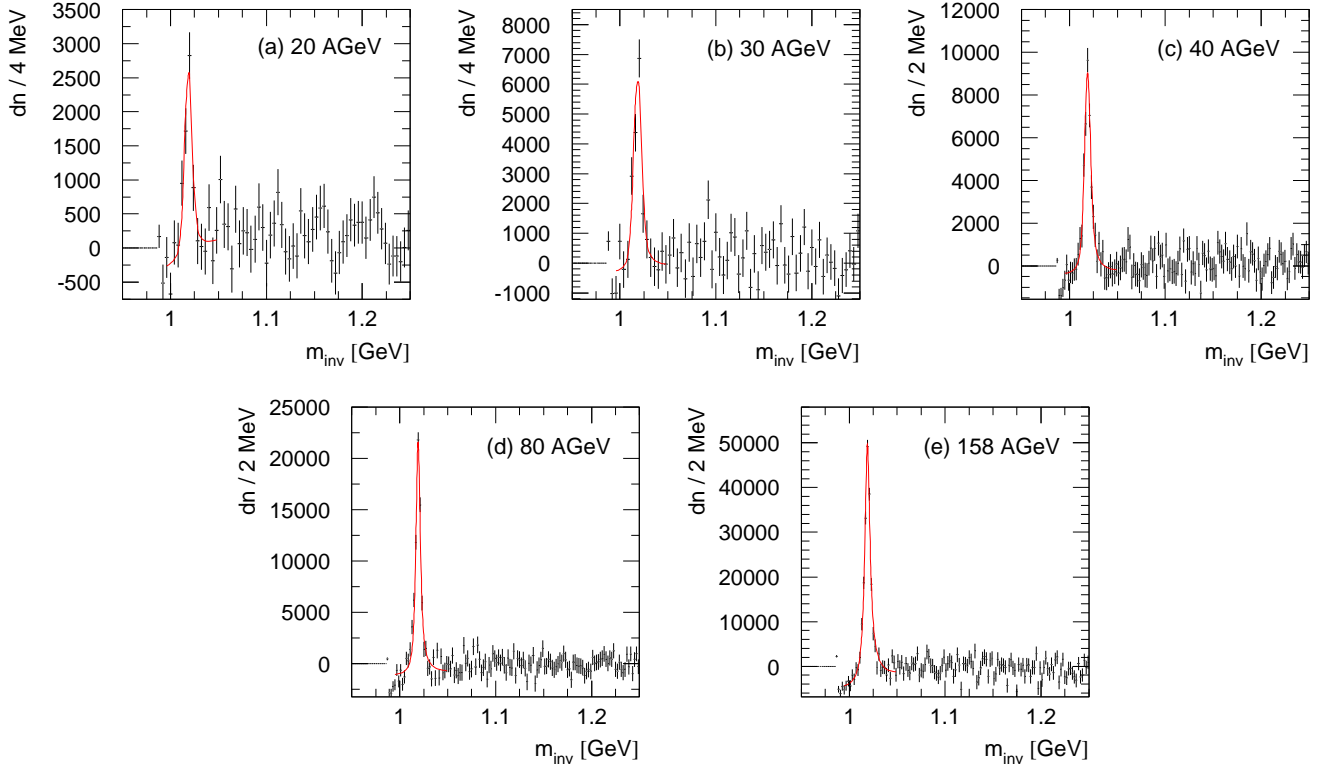}
\end{center}
\caption{(Color online) $K^+ K^-$ invariant-mass spectra after subtraction of the 
combinatorial background in the forward rapidity hemisphere for the
five different beam momenta. The full lines show the Breit-Wigner fits
to the signals as described in the text. The bin size is 4~MeV for
20$A$ and 30$A$~GeV and 2~MeV for the other beam energies.}
\label{fig:signals}
\end{figure*}

The $\phi$ signal was obtained by calculating the invariant mass of all
combinations of positive and negative kaon candidates in one event.
To reconstruct the combinatorial background of uncorrelated pairs,
candidates from different events were combined.
The mixed-event spectrum was subtracted from the same-event spectrum
after normalization to the same number of pairs~\cite{drijard1984}. 
Figure~\ref{fig:signals} shows the background-subtracted invariant-mass 
spectra in the total forward acceptance for different collision energies.
In all cases, clear signals are observed at the expected position.

\begin{figure*}
\includegraphics[width=0.45\linewidth]{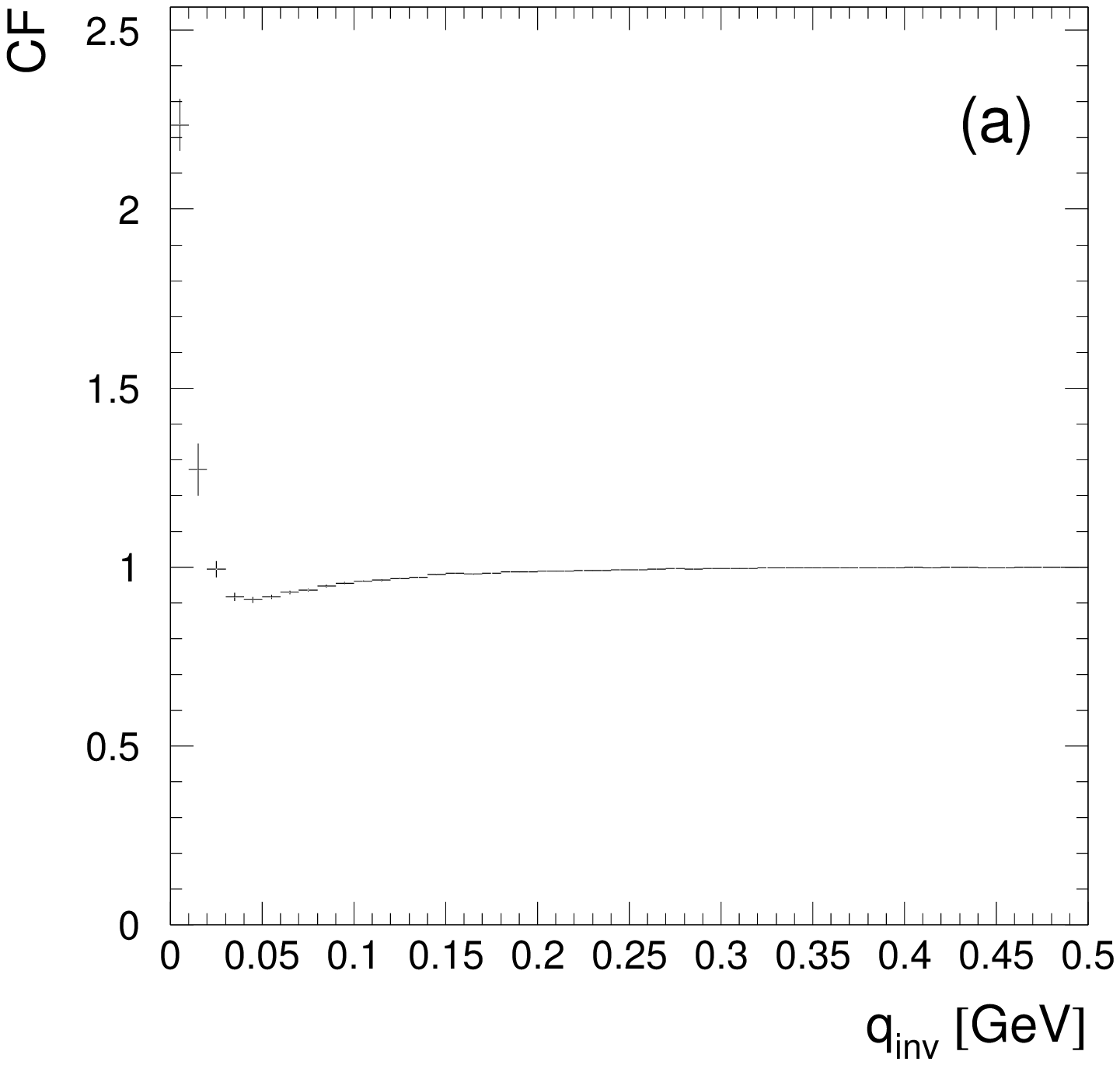}
\includegraphics[width=0.45\linewidth]{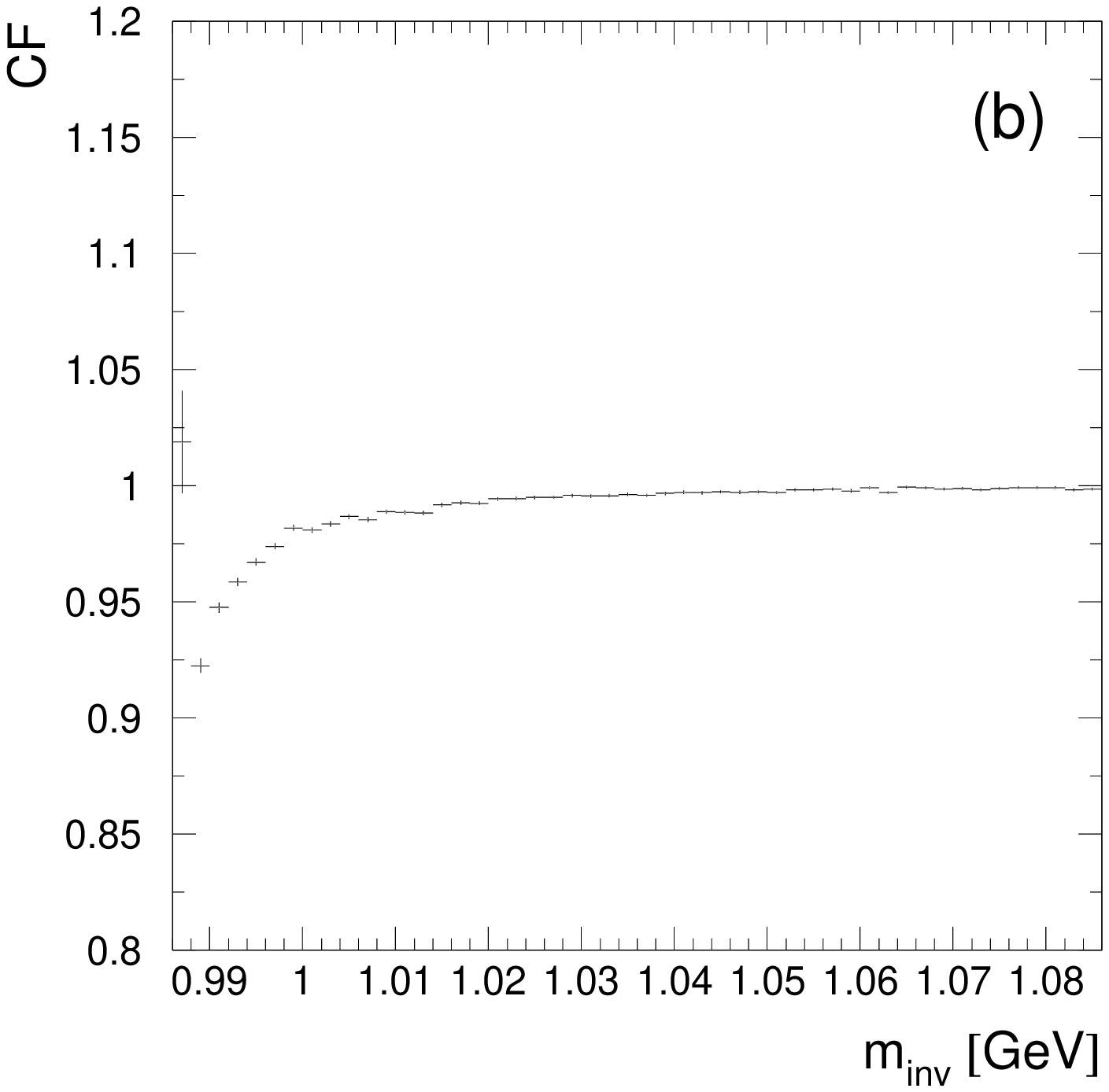}
\caption{$K^+K^-$ correlation function close to threshold
in (a) $q_{\rm inv}$ and (b) $m_{\rm inv}$~\cite{lednicki}. }
\label{fig:fsikk}
\end{figure*}

While the subtracted spectrum is flat on the right side of the signal,
a depletion is observed between the peak and the threshold. As a possible
source of this undershoot, the correlation of kaons stemming from
different $\phi$ mesons has been discussed in Ref.~\cite{drijard1984}. 
In our case, it was shown by simulation that this effect is small thanks to 
the large acceptance of the NA49 MTPCs. Another possible source of the 
distortion is the reflection of other resonances, 
e.g., $\Delta^0 \to \pi^-p$, into the
$K^+K^-$ spectrum by misidentification of pions and protons, as discussed
in detail in Ref.~\cite{friese1999}. This effect was shown to be present in
our previous analysis of another data set~\cite{na492000}, 
where the $dE/dx$ resolution was significantly worse. However, all 
such resonances would distort the spectrum over a broad range above 
threshold, which can be excluded by the observed flatness at higher 
masses. This conclusion is further strengthened by the observation that 
the depletion does not vanish when applying a stricter $dE/dx$ cut on 
the kaons. Hence, the undershoot is likely to originate from a true 
correlation of kaon pairs. 
Simulations show that it can be explained by final state strong
interaction of kaons~\cite{lednicki}.
This is demonstrated in Fig.~\ref{fig:fsikk} by
showing the $K^+K^-$ correlation function in 
$q_{\rm inv} = \sqrt{ (\vec p_1  - \vec p_2)^2 - (E_1 - E_2)^2 }$
[Fig.~\ref{fig:fsikk}(a)] and 
in $m_{\rm inv}$ [Fig.~\ref{fig:fsikk}(b)]. While the repulsive interaction causes a depletion 
in $m_{\rm inv}$, the stronger attractive Coulomb effect is squeezed 
into 0.8~MeV above threshold and is thus hardly seen. In combination
with the steeply rising unsubtracted $m_{\rm inv}$ distribution, this
depletion can easily account for the deficit observed in the subtracted spectrum.

To correct this effect quantitatively by simulation is difficult and would
moreover be model dependent. As the narrow signal is easily distinguished from
the broad residual background, we accounted for the depletion by fitting 
a straight line in the vicinity of the peak. For the description of the 
signal itself, we used a relativistic $p$-wave Breit-Wigner 
distribution~\cite{jackson1964} of the form
\begin{equation}
\frac{dN}{dm} \propto \frac { m \Gamma(m) }
{ (m^2-m_0^2)^2 + m_0^2\Gamma^2(m) }
\end{equation}
with the mass-dependent width
\begin{equation}
\Gamma(m) = 2 \, \Gamma_0 \left( \frac {q} {q_0} \right) ^3 \,
\frac {q_0^2} {q^2+q_0^2} \, ,
\end{equation}
where $q:=\sqrt{ \frac{1}{4} m^2 - m_K^2}$ and 
$q_0:= \sqrt{ \frac{1}{4} m_0^2 - m_K^2}$.
This distribution was folded with a Gaussian representing the
invariant-mass resolution $\sigma_m$ of the spectrometer. 
Since in general, mass resolution and width cannot be determined 
separately, we fixed the width to its book value 
$\Gamma_0 = 4.26 \, {\rm MeV}$~\cite{pdg2004}, 
leaving $m_0$, $\sigma_m$, a normalization and two parameters for the
linear background as free parameters for the fit, which was performed
in the mass range 994--1050~MeV. It was checked by
simulations that this procedure gives the correct values for position, 
width and integral of the distribution. As Fig.~\ref{fig:signals} 
demonstrates, the fit gives a good description of the signal. 
The numerical values of the fitted parameters are listed in 
Table~\ref{tab:signals}.

To obtain longitudinal and transverse spectra, the signal was extracted
in rapidity and in $p_t$ bins, respectively,
in the same way as in the total acceptance. 
Generally, the limited statistics prevented a simultaneous division into
$y$-$p_t$ bins. Thus, transverse momentum spectra could only be derived
averaged over rapidity.
To reduce the number of free
fit parameters, $m_0$ and $\sigma_m$ were fixed for the fits
in the phase space bins to the values obtained from the
signal in the total acceptance. 
For the 158$A$~GeV data set, where the statistics in the signal
allowed to do so, we checked that leaving these parameters free did not
significantly alter the results. In particular, no significant
dependence of $m_0$ or $\sigma_m$ on rapidity or $p_t$ was observed.

Since the straight-line background is only an approximation for the
residual background in the vicinity of the signal, the stability of the fit against
the variation of the fit region was checked. The parameters $m_0$
and $\sigma_m$ show no significant dependence. The variation of the 
normalization constants, which determine the fit integral, is in
all bins far below the statistical error returned by the fit
procedure. We conclude that the latter properly takes into account
the possible variations of the baseline.

The raw yields in the phase-space bins were obtained by integrating
the fit function from threshold up to 
$m_0 + 30 \Gamma_0 \approx 1.148$~MeV. 
This mass cutoff is somehow arbitrary; the corresponding integral
varies by about 3\% for cutoff values from $m_0 + 10 \Gamma_0$
to infinity.
We take this as a systematic uncertainty due to the mass cutoff.
Using alternatively a (analytically integrable) nonrelativistic
Lorentz distribution for the fit does not change the integral 
by more than 1\%. 

\subsection{Geometrical acceptance}
\begin{figure*}
\includegraphics[width=0.45\linewidth]{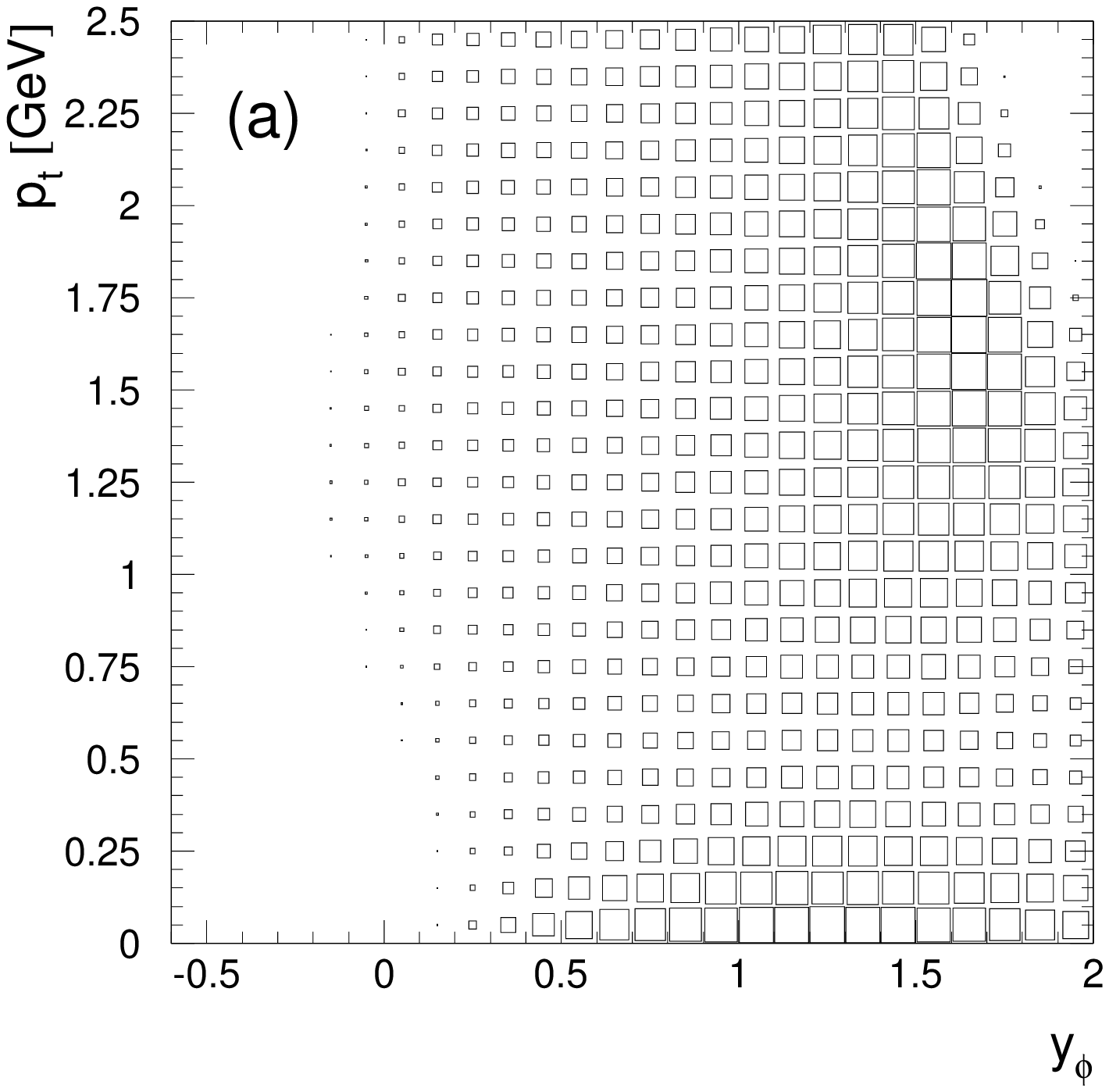}
\includegraphics[width=0.45\linewidth]{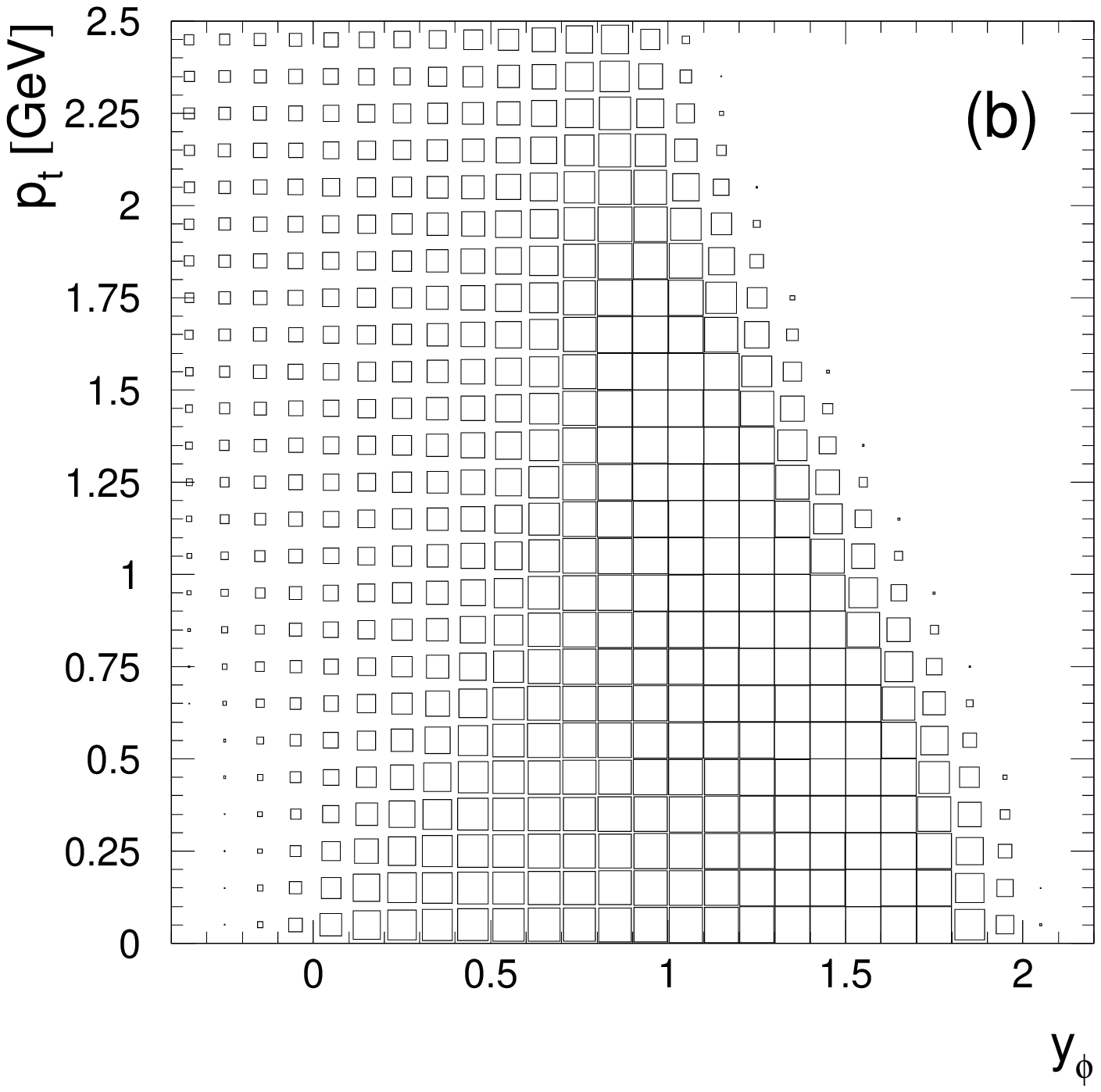}
\caption{Geometrical acceptance probability for $\phi \to K^+K^-$ including kaon decay in
flight for (a) 20$A$~GeV and (b) 158$A$~GeV. }
\label{fig:acceptance}
\end{figure*}

The geometrical acceptance of the NA49 detector for the decay  
$\phi \rightarrow K^+ K^-$ was obtained double-differentially in $y$ 
and $p_t$ (integrated over azimuth) by {\sc geant} simulations of the 
$\phi$ decay  including in-flight decay of the 
kaon daughters, assuming an azimuthally flat $\phi$ emission and isotropic 
decay. The resulting
acceptance is shown in Fig.~\ref{fig:acceptance} for 20$A$ and 158$A$~GeV. 
While the upper momentum limit for the daughter candidates restricts 
the acceptance
at forward rapidity for the top SPS energy, at lower beam energies there is
lack of acceptance near midrapidity because of the lower momentum limit
for the daughter tracks and the increased losses due to in-flight decay 
for low-momentum kaons.

As the acceptance is a function of $y$ and $p_t$, the proper correction
factor for a given extended phase-space bin (integrated either over 
$y$ or $p_t$) as used in the analysis is the mean acceptance
\begin{equation}
\overline{a}_S = \frac
{ \int \limits _S dy dp_t \, a(y,p_t) \, f(y,p_t) }
{ \int \limits _S dy dp_t \, f(y,p_t) } \, ,
\label{eq:acceptance}
\end{equation}
where $S$ denotes the region in the $y,p_T$ plane, $a(y,pt)$ the
acceptance probability averaged over the azimuthal angle, and 
$f(y,p_t)$ the differential $\phi$ meson yield. For the rapidity
distributions, the differential yields have in addition to be extrapolated
to the full $p_t$ range. The extrapolation factor, however, is small
($<$5\%) due to the large $p_t$ range covered by the experiment.

Both the acceptance correction and the extrapolation to full $p_t$
require the knowledge of the $y$ and $p_t$ dependence of $\phi$ meson yields, which 
leads to an iterative procedure (see Sec.~\ref{sec:spectra}).

\subsection{Spectra and yields}
\label{sec:spectra}

Apart from the differential acceptance correction, the raw yields
obtained from the fit to the invariant-mass spectra were corrected
for the branching ratio $\phi \rightarrow K^+ K^-$ 
(49.1\%) and the efficiency of kaon $dE/dx$ selection (75\% for the pair), and
normalized to the number of collisions. These global correction factors 
are common for all bins in phase space and for all beam energies.

The transverse spectra are fitted by the thermal ansatz
\begin{equation}
\frac {dn} {dp_t} \propto p_t \, e^{ -m_t / T } \, ,
\label{eq:ptthermal}
\end{equation}
where the transverse mass $m_t = \sqrt{m_0^2 + p_t^2}$. The distributions
in rapidity were parametrized by a single Gaussian

\begin{equation}
\frac {dn} {dy} \propto e^{ - \frac {y^2} {2 \sigma_y^2} } \, .
\label{eq:rapgauss}
\end{equation}

As the parameters $T$ and $\sigma_y$ must be obtained by
the analysis itself, an iterative procedure was employed.
Starting from some reasonable parameter values, the
acceptance correction was calculated according to Eq.~(\ref{eq:acceptance}),
assuming factorization of the emission function $f(y, p_t)$ into
the transverse and longitudinal distributions~(\ref{eq:ptthermal})
and~(\ref{eq:rapgauss}), i.e., independence of $T$ on rapidity.
The corrected yields in the $p_t$ and $y$ bins were then fitted
with the distributions~(\ref{eq:ptthermal}) and (\ref{eq:rapgauss}),
respectively, obtaining new values for $T$ and $\sigma_y$ which
serve as input for the next iteration.  Convergence of the method was 
reached after three to five steps. It was checked that the final results
do not depend on the choice of start values for the parameters.

After the final step of the iteration, the yields in full phase space
were obtained by summing up the measured yields in the rapidity 
distributions and numerically extrapolating Eq.~(\ref{eq:rapgauss})
to the full rapidity range. In a similar way, the quantities 
$\langle p_t \rangle$, $\langle m_t \rangle$, and $\sigma_y$ 
were determined. The midrapidity yield $dn/dy$ was obtained 
directly from the fit function.

As demonstrated later in Fig.~\ref{fig:yspectra}, the Gaussian
parametrization gives a satisfactory description of the 
rapidity distribution for all data sets. However, because of the lack
of midrapidity data points at the lower beam energies, an 
ambiguity for the extrapolation to full phase space arises.
To check the sensitivity of the results to the assumed shape
of the rapidity distribution, we alternatively parametrized
the latter by the sum of two Gaussian functions displaced
symmetrically around midrapidity by a shift $a$:

\begin{equation}
\frac {dn} {dy} \propto e^{ - \frac {(y-a)^2} {2 \sigma_y^2} }
+ e^{ - \frac {(y+a)^2} {2 \sigma_y^2} } \, .
\label{eq:rap2gauss}
\end{equation}

The width of this distribution will be characterized by its rms value.
Total yield $\langle \phi \rangle$, midrapidity yield $dn_\phi/dy$
and ${\rm rms}_y$ were calculated
for both parametrizations (\ref{eq:rapgauss}) and (\ref{eq:rap2gauss}).
The final values listed in Tables~\ref{tab:yspectra} and~\ref{tab:yields} were
calculated as the mean of the results of the two methods; 
their differences enter the systematic errors.  

\subsection{Statistical and systematic errors}
\label{sec:errors}

Statistical errors in the raw differential $\phi$ meson yields originate 
from the statistical bin-by-bin errors in the same-event and 
mixed-event invariant mass spectra, which 
were found to be in good approximation Poissonian 
and uncorrelated between mass bins. Then, the statistical errors in 
the event-mix subtracted invariant-mass spectrum was calculated 
as~\cite{drijard1984}
\begin{equation}
\sigma_i^2 = n_{0,i} + k^2 n_{{\rm em},i} \, ,
\end{equation}
where $n_{0,i}$ is the number of entries in mass bin $i$ in the 
same-event spectrum, $n_{{\rm em},i}$ the same number in the mixed-event 
spectrum, and $k$ the normalization constant for the event mix. 
These errors were propagated toward the raw differential yields 
by the least-squares fit of the Breit-Wigner distribution to the signal 
peak.

The acceptance calculation was performed with sufficiently high statistics such
that the relative statistical error of the differential acceptance is
below 1\% and thus far below the uncertainty in the raw yields 
over the entire $y$, $p_t$ region used for the analysis. 
Finally, the errors in the acceptance-corrected differential yields are
propagated through the least-square fits to the spectra
to obtain the statistical uncertainties in the spectral parameters and
the integrated quantities.

Systematic uncertainties in the uncorrected yields arise from the
approximation of the residual background in the invariant-mass spectra
as a straight line. This approximation is only valid in a limited
mass range around the signal peak. Thus, the stability of the results
of the Breit-Wigner fit against the variation of the fit range was
checked. We found no significant dependence of the parameters $m_0$
and $\sigma_m$; the variation of the normalization constant, determining 
the fit integral, was in all $y$ and $p_t$ bins found to be smaller
than the statistical error. 

Another source of systematic error arises from the $dE/dx$ selection 
of kaon candidates. Uncertainties in the parametrization of the mean kaon 
$dE/dx$ and the resolution result in systematic deviations of the 
efficiency correction from its true value. To estimate
this error, the analysis was repeated for different widths of the
$dE/dx$ selection window around the kaon expectation value, applying the
respective efficiency correction. This error was found to be the
dominating one; for most raw yields, it is comparable to or slightly 
larger than the statistical one.

Imperfect detector description in the simulation leads to systematic
uncertainties in the acceptance correction. To reduce possible errors,
the analysis was restricted to phase-space regions where the acceptance
is above 1\%. The remaining error was estimated by repeating the analysis 
with varying acceptance conditions (minimal track length in the MTPCs).
It was in all cases found to be much smaller than the error originating
from the kaon selection by $dE/dx$.

As the spectral parameters enter the acceptance correction through
Eq.~(\ref{eq:acceptance}), their uncertainties add to the systematic
errors of the corrected yields. This was accounted for by determining
the range of acceptance values allowed by the errors in $T$ and 
$\sigma_y$. In addition, for the rapidity bins close to beam rapidity,
a possible deviation of the slope parameter by $50$~MeV from its
averaged value was taken into account in the acceptance correction.
The resulting error, however, is small thanks to the large and
approximately uniform $p_t$ acceptance.

The systematic errors in the corrected differential yields were
assumed to be independent and added in quadrature. They were
propagated to the respective errors in the spectral parameters
by repeating the fit of Eqs.~(\ref{eq:ptthermal})--(\ref{eq:rap2gauss})
with statistical
and systematic errors added and comparing the resulting errors
to those obtained from the fit with statistical errors only.

For the determination of the averaged quantities 
$\langle \phi \rangle$, $\langle p_r \rangle$,
$\langle m_t \rangle$, and ${\rm rms}_y$, the summation of the measured
differential yields as well as extrapolation to full phase
space are required. The systematic errors of these observables
were determined from the errors of the differential yields and
the uncertainties in the spectral shapes.

\section{Results}

\subsection{Line shape}

Table~\ref{tab:signals} summarizes the parameters obtained from the
invariant-mass signals in the total acceptance. The signal quality
decreases when going to lower beam energy because of both the reduced 
$\phi$ meson yield and the reduced acceptance due to the increased in-flight
decay probability for the daughter kaons. At all five energies, the fitted peak
position is slightly below the literature value of 1019.43~MeV~\cite{pdg2004}.
We investigated the effect of an error in the normalization of the  magnetic 
field used for momentum determination in the reconstruction chain 
and found that a bias of 1\% in the magnetic field is needed to explain the observed shift.
This is slightly above the momentum scale uncertainty deduced from a precision 
study of the
$K^0_s$ signal. We thus cannot exclude that the deviation of the peak position
is due to experimental effects.

\begin{table*}
\caption{\label{tab:signals}Approximate number of detected $\phi$ mesons $S$, background-to-signal
ratio $B/S$, signal-to-noise ratio SNR, position of the signal peak $m_0$, and 
invariant-mass resolution $\sigma_m$. The latter two were obtained by a Breit-Wigner
fit to the signal peak (see text). The width was fixed to its literature
value 4.26 MeV. $S$ and $B$ were calculated in a window of $\pm$ 4 MeV around
the peak. The quoted errors are statistical only.}
\begin{center}
\begin{ruledtabular}
\begin{tabular}{crrrcc}
$p_{\rm beam}$  &  $S$  &  $B/S$  &  SNR  &  $m_0$  &  $\sigma_m$ \\
($A~$GeV)        &       &         &       &  (MeV)  &   (MeV)  \\
\hline
 20  &    6 500   &  70  &   9.4  &  $1018.8 \pm 0.6$  &  $2.6 \pm 0.9$ \\
 30  &   16 500   & 104  &  12.5  &  $1018.4 \pm 0.5$  &  $2.5 \pm 1.3$ \\
 40  &   37 000   &  53  &  26.2  &  $1018.9 \pm 0.2$  &  $2.1 \pm 0.3$ \\
 80  &   55 000   &  30  &  42.5  &  $1019.1 \pm 0.1$  &  $1.1 \pm 0.1$ \\
158  &  180 000   &  72  &  49.4  &  $1019.0 \pm 0.1$  &  $1.8 \pm 0.1$ \\   
\end{tabular}
\end{ruledtabular}
\end{center}
\end{table*} 

The widths of the mass peaks obtained from the fits are consistent with those
obtained from a full detector simulation and reconstruction. Their slight
increase toward lower beam energies can be understood as the increasing
influence of multiple scattering on lower momentum tracks. For the
signal at 158$A$~GeV, we fitted simultaneously width and mass resolution
and obtained $\Gamma_0 = (4.41 \pm 0.61) \, \rm{MeV}$, 
$\sigma_m = (1.81 \pm 0.26) \, \rm{MeV}$, i.e., no deviation from the
free-particle width. Thus, within experimental uncertainties, we do
not observe indications for a mass shift or a broadening of the $\phi$
meson.

The observation that the mass and width of the $\phi$ meson agree
with the Particle Data Group values is in line with the results of
AGS and RHIC experiments~\cite{back2004,adler2003,adams2005,adler2005}.
It should be noted that because of the long lifetime of the $\phi$ meson
($\tau = 46$~fm), only a fraction decays inside the fireball. 
Thus, only a part of the $\phi$ mesons can be expected to be
influenced by the surrounding medium.

\begin{table*}
\caption{\label{tab:data}Differential $\phi$ meson yields in the $p_t$ (left) and
$y$  (right) distributions. Data in the $p_t$ bins are integrated over the
rapidity ranges given in Table~\ref{tab:ptspectra}. The errors are statistical.}
\begin{center}
\begin{ruledtabular}
\begin{tabular}{cccc}
$p_t$ (GeV)  &  $dn / (dy dp_t)$ (GeV$^{-1})$  &   y  &  $dn/dy$  \\
\hline
\multicolumn{4}{c}{$E_{beam}$ = 20$A$~GeV} \\
0.0--0.4  &  $0.382 \pm 0.074$ &  0.2--0.6 & $1.043 \pm 0.250$ \\
0.4--0.8  &  $0.528 \pm 0.097$ &  0.6--1.0 & $0.536 \pm 0.077$ \\
0.8--1.2  &  $0.257 \pm 0.054$ &  1.0--1.4 & $0.159 \pm 0.033$ \\
1.2--1.6  &  $0.079 \pm 0.030$ &  1.4--1.8 & $0.032 \pm 0.017$ \\
1.6--2.0  &  $0.033 \pm 0.015$ &  \\
 & & & \\

\multicolumn{4}{c}{$E_{beam}$ = 30$A$~GeV} \\
0.0--0.3  &  $0.231 \pm 0.051$ &   0.3--0.6 & $0.735 \pm 0.194$ \\
0.3--0.6  &  $0.578 \pm 0.079$ &   0.6--0.9 & $0.651 \pm 0.090$ \\
0.9--1.2  &  $0.386 \pm 0.079$ &   0.9--1.2 & $0.456 \pm 0.052$ \\
1.2--1.5  &  $0.257 \pm 0.050$ &   1.2--1.5 & $0.193 \pm 0.036$ \\
1.5--1.8  &  $0.070 \pm 0.019$ &   1.5--1.8 & $0.097 \pm 0.029$ \\
& & & \\

\multicolumn{4}{c}{$E_{beam}$ = 40$A$~GeV} \\
0.0--0.2  &  $0.185 \pm 0.035$ &  0.3--0.6 & $1.067 \pm 0.108$ \\
0.2--0.4  &  $0.668 \pm 0.052$ &  0.6--0.9 & $0.756 \pm 0.059$ \\
0.4--0.6  &  $0.780 \pm 0.064$ &  0.9--1.2 & $0.611 \pm 0.038$ \\
0.6--0.8  &  $0.625 \pm 0.075$ &  1.2--1.5 & $0.348 \pm 0.028$ \\
0.8--1.0  &  $0.569 \pm 0.073$ &  1.5--1.8 & $0.188 \pm 0.023$ \\
1.0--1.2  &  $0.413 \pm 0.059$ &  \\
1.2--1.4  &  $0.275 \pm 0.040$ &  \\
1.4--1.6  &  $0.081 \pm 0.028$ &  \\
1.6--1.8  &  $0.086 \pm 0.019$ &  \\
1.8--2.0  &  $0.057 \pm 0.014$ &  \\
& & & \\

\multicolumn{4}{c}{$E_{beam}$ = 80$A$~GeV} \\
0.0--0.2  &  $0.337 \pm 0.031$ &  -0.3--0.0 & $1.591 \pm 0.304$ \\
0.2--0.4  &  $0.886 \pm 0.051$ &   0.0--0.3 & $1.474 \pm 0.138$ \\
0.4--0.6  &  $1.148 \pm 0.057$ &   0.3--0.6 & $1.258 \pm 0.086$ \\
0.6--0.8  &  $0.996 \pm 0.056$ &   0.6--0.9 & $1.351 \pm 0.062$ \\
0.8--1.0  &  $0.861 \pm 0.052$ &   0.9--1.2 & $1.041 \pm 0.049$ \\
1.0--1.2  &  $0.517 \pm 0.048$ &   1.2--1.5 & $0.718 \pm 0.043$ \\
1.2--1.4  &  $0.344 \pm 0.045$ &   1.5--1.8 & $0.408 \pm 0.037$ \\
1.4--1.6  &  $0.173 \pm 0.040$ &   1.8--2.1 & $0.197 \pm 0.040$ \\
& & & \\

\multicolumn{4}{c}{$E_{beam}$ = 158$A$~GeV} \\
0.0--0.2  &  $0.582 \pm 0.053$ &   0.0--0.2 & $2.557 \pm 0.166$ \\
0.2--0.4  &  $1.275 \pm 0.086$ &   0.2--0.4 & $2.386 \pm 0.121$ \\
0.4--0.6  &  $1.924 \pm 0.098$ &   0.4--0.6 & $2.229 \pm 0.098$ \\
0.6--0.8  &  $2.016 \pm 0.099$ &   0.6--0.8 & $2.202 \pm 0.089$ \\
0.8--1.0  &  $1.778 \pm 0.092$ &   0.8--1.0 & $1.974 \pm 0.090$ \\
1.0--1.2  &  $1.339 \pm 0.080$ &   1.0--1.2 & $1.816 \pm 0.094$ \\
1.2--1.4  &  $0.956 \pm 0.067$ &   1.2--1.4 & $1.636 \pm 0.105$ \\
1.4--1.6  &  $0.567 \pm 0.055$ &   1.4--1.6 & $1.528 \pm 0.126$ \\
1.6--1.8  &  $0.370 \pm 0.044$ &   1.6--1.8 & $1.125 \pm 0.171$ \\
1.8--2.0  &  $0.200 \pm 0.034$ &    \\
\end{tabular}
\end{ruledtabular}
\end{center}
\end{table*} 

\subsection{Transverse momentum spectra}

\begin{figure*}
\includegraphics[width=\linewidth]{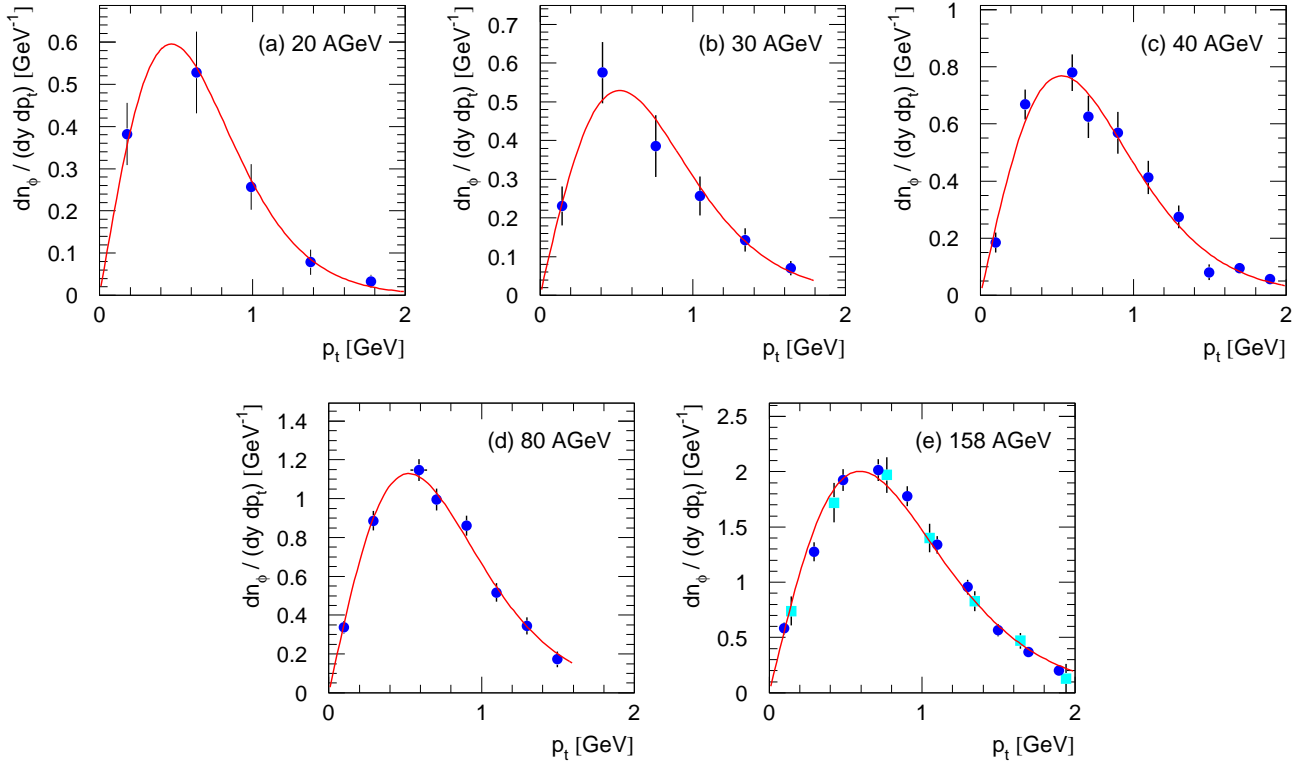}
\caption{(Color online) $\phi$ transverse momentum spectra integrated over the rapidity
intervals given in Table~\ref{tab:ptspectra}. The full lines show the
fits of thermal distributions (\ref{eq:ptthermal}). The squared symbols denote
previously published results~\cite{na492000}. Only statistical
errors are shown.}
\label{fig:ptspectra}
\end{figure*}

\begin{figure*}
\includegraphics[width=0.45\linewidth]{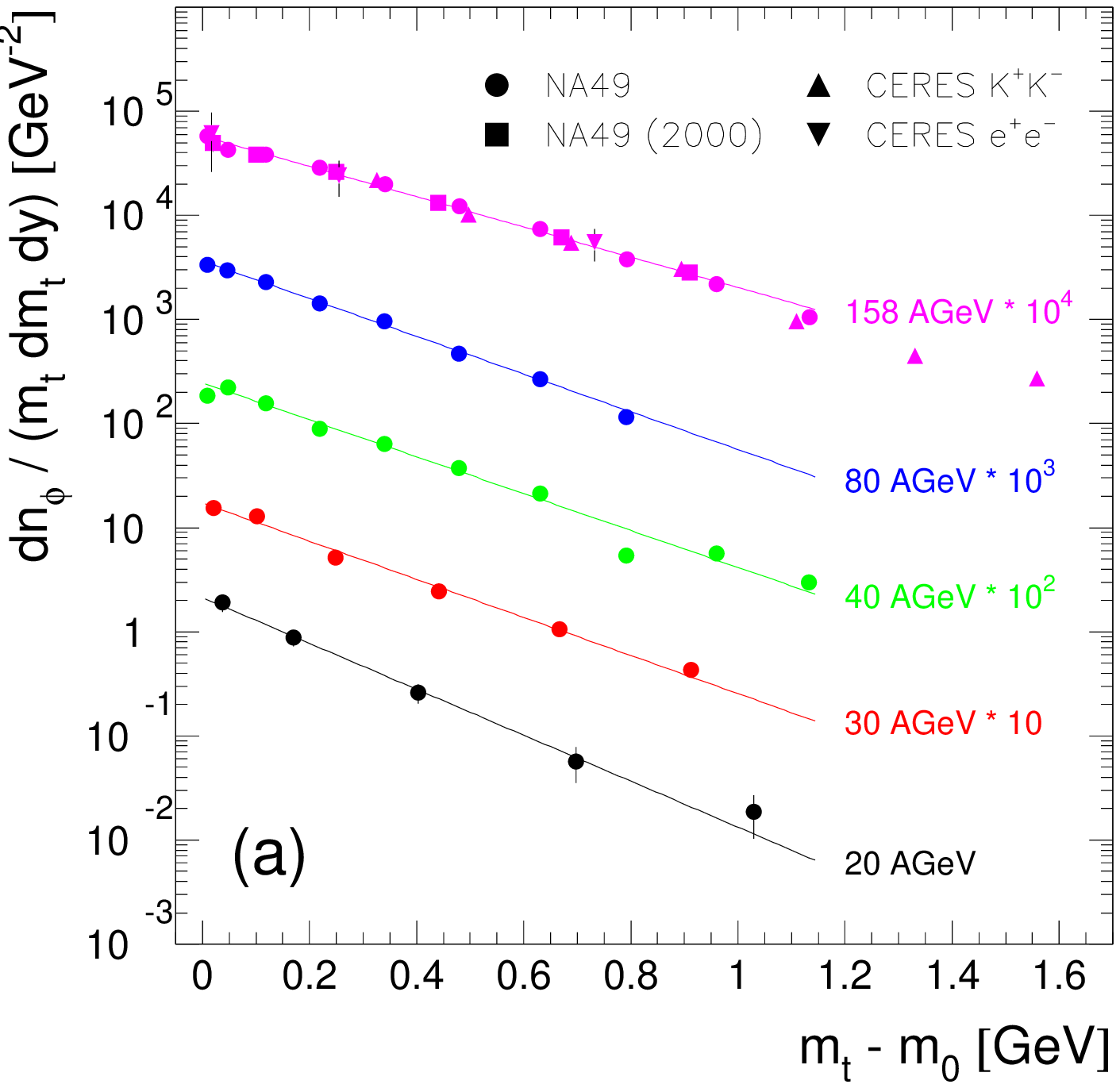}
\includegraphics[width=0.45\linewidth]{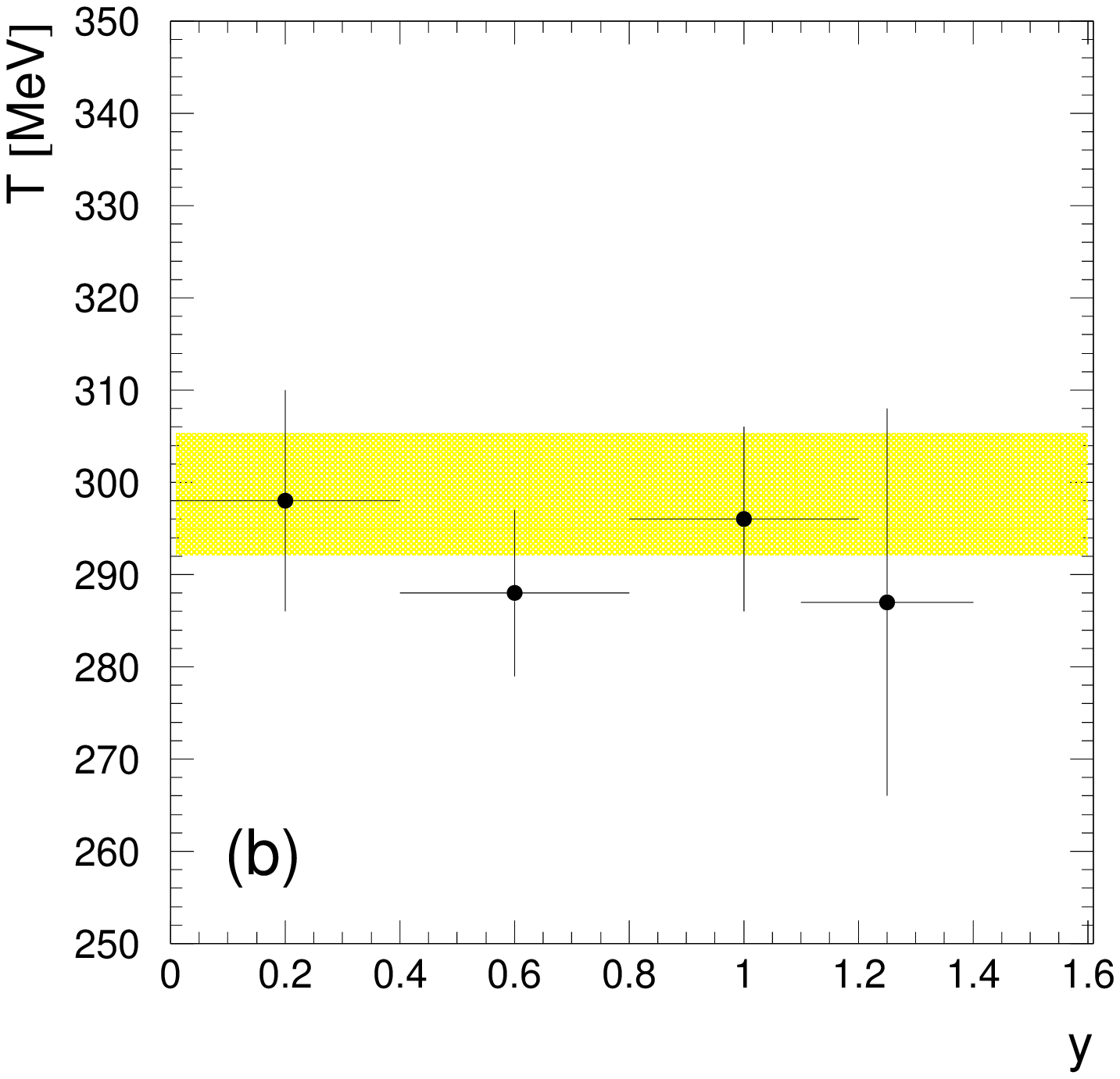}
\caption{(Color online) (a) $\phi$ transverse mass spectra integrated over the rapidity
intervals given in Table~\ref{tab:ptspectra}. The exponential fits
indicated by the full lines correspond to the fits shown in
Fig.~\ref{fig:ptspectra}. The spectra for different beam energies are scaled
for better visibility. Only statistical errors are shown. The data at 158$A$~GeV 
are compared with previously published results
of NA49~\cite{na492000} and CERES~\cite{adamova2006}.
(b) Slope parameter as function of rapidity at 158$A$~GeV. 
The values agree within errors with that obtained from the $y$-integrated
$p_t$ spectrum, the latter indicated with its standard deviation by the shaded bar.}
\label{fig:mtspectra}
\end{figure*}

The transverse momentum spectra obtained for the five beam energies are
shown in Fig.~\ref{fig:ptspectra}; numerical data are given in 
Table~\ref{tab:data}. In all cases, the thermal distribution
(\ref{eq:ptthermal}) gives a good description of the data; the fit
parameters are summarized in Table~\ref{tab:ptspectra}. 
At top SPS energy with the best signal quality, a modest deviation 
from the fit function is indicated  by the $\chi^2/\rm{ndf}$ of 1.5. 
A slight curvature of the transverse mass spectrum at this energy, as 
expected from a hydrodynamical expansion scenario, is visible for this 
energy in Fig.~\ref{fig:mtspectra}(a). For the other energies, 
no deviations from pure exponential behavior can be seen within 
the experimental uncertainties.

The transverse momentum spectrum can be also characterized by its first
moment or the average transverse mass. These parameters were calculated
from the measured data points and extrapolated to full $p_t$ using
the exponential fit function. As the extrapolation contributes only marginally because of the
large $p_t$ coverage, $\langle p_t \rangle$ and 
$\langle m_t \rangle - m_0$ are largely independent of the spectral shape. 
Their values are also listed in Table~\ref{tab:ptspectra}.

\begin{table*}
\caption{\label{tab:ptspectra}Rapidity range (in c.m. system), $p_t$ range, slope parameter $T$, 
$\chi^2$ per degree of freedom, average $p_t$, and average $m_t$
for the transverse momentum spectra. $T$ and $\chi^2$ are results
from the fit of Eq.~(\ref{eq:ptthermal}) to the spectrum; 
$\langle p_t \rangle$ and $\langle m_t \rangle - m_0$ were
obtained by summation over the data points and extrapolation to
full $p_t$ using the fit function. The first error is statistical,
the second one systematic.}\begin{center}
\begin{ruledtabular}
\begin{tabular}{ccccccc}
$p_{\rm beam}$ ($A$~GeV) & $y$ range & $p_t$ range (GeV) & $T$ (MeV) 
& $\chi^2 / \rm{ndf}$ & $\langle p_t \rangle$ (MeV) 
& $\langle m_t \rangle - m_0$ [MeV]\\
\hline
 20  & 0.0--1.8 & 0.0--2.0  & $196.8 \pm 19.5 \pm 20.2 $ & 1.06/3 
     & $650.9 \pm 34.2 \pm 40.2$ & $229.5 \pm 20.1 \pm 23.6$ \\
 30  & 0.0--1.8 & 0.0--1.8  & $237.4 \pm 17.8 \pm 22.9 $ & 2.03/4
     & $738.9 \pm 28.3 \pm 46.3$ & $284.6 \pm 17.3 \pm 28.4$ \\
 40  & 0.0--1.5 & 0.0--2.0  & $244.6 \pm  9.0 \pm  5.8 $ & 12.42/8
     & $763.4 \pm 15.8 \pm 14.3$ & $297.8 \pm 10.0 \pm 9.2$ \\
 80  & 0.0--1.7 & 0.0--1.6  & $239.8 \pm  8.3 \pm 10.9 $ & 3.48/6
     & $756.4 \pm 11.5 \pm 22.5$ & $292.6 \pm 7.6 \pm 15.3$ \\
158  & 0.0--1.0 & 0.0--2.0  & $298.7 \pm  6.6 \pm 10.6 $ & 12.06/8
     & $883.5 \pm 9.9 \pm 21.3$ & $378.3 \pm 6.7 \pm 15.2$ \\   
\end{tabular}
\end{ruledtabular}
\end{center}
\end{table*} 

The assumption of the slope parameter being independent of $y$ could
be checked for 158$A$~GeV, where statistics allowed us to extract
transverse spectra in four different rapidity bins. The resulting slope
parameters are shown in Fig.~\ref{fig:mtspectra}(b). Within the
measured rapidity range, we observe no significant change of the
slope parameter with $y$. Using the $y$-dependent slope parameters
for correcting the rapidity distribution had no sizable effect
on the results.

The spectrum obtained for 158$A$~GeV agrees with that
from an earlier publication~\cite{na492000} of the NA49 experiment, 
which was based on the analysis of an older data set at the same beam 
energy. For comparison, the previously published data are shown by the square
symbols in Figs.~\ref{fig:ptspectra}(e) and~\ref{fig:mtspectra}(a).
There is agreement with the results of the CERES experiment
in both decay channels
$\phi \rightarrow K^+ K^-$ and 
$\phi \rightarrow e^+ e^-$~\cite{adamova2006}, as also demonstrated
in Fig.~\ref{fig:mtspectra}(a).
The data disagree with the spectrum
measured by the NA50 experiment in the di-muon decay channel
$\phi \rightarrow \mu^+ \mu^-$, where a significantly smaller slope
was obtained~\cite{alessandro2003}. 

\subsection{Rapidity distributions and yields}

\begin{figure*}
\includegraphics[width=\linewidth]{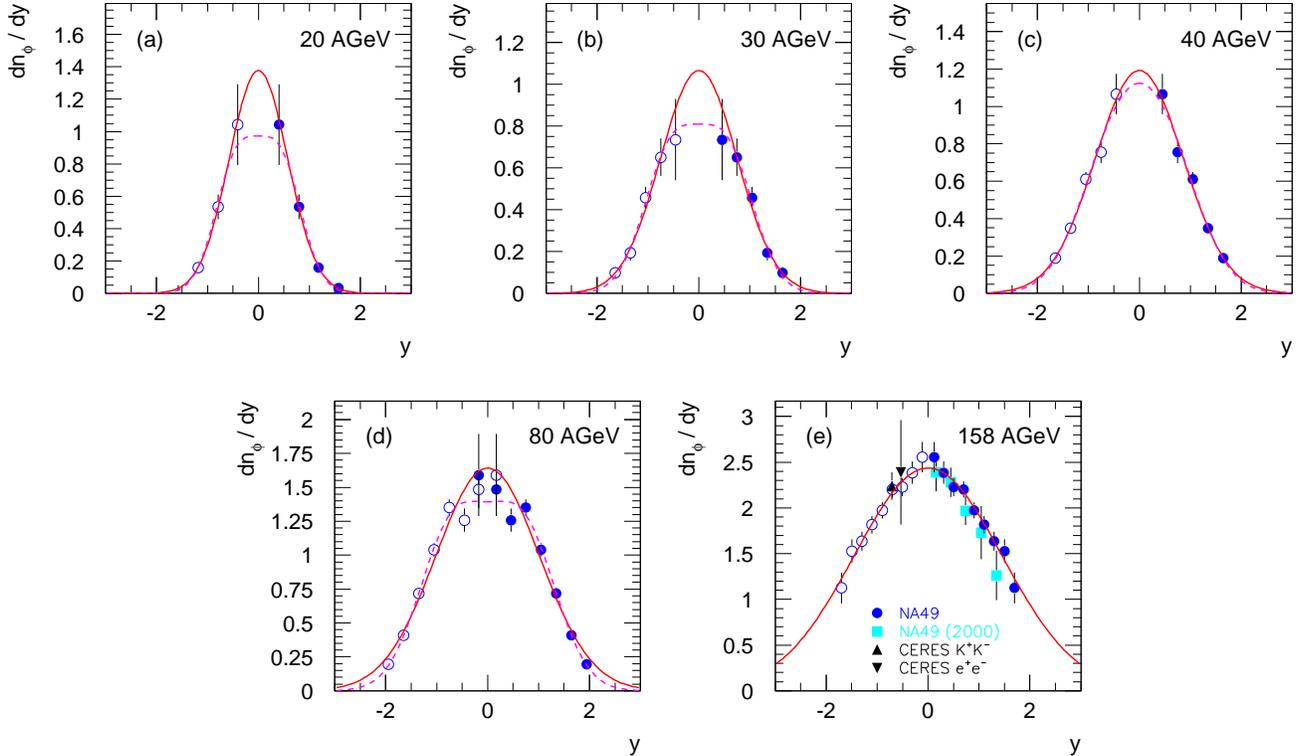}
\caption{(Color online) $\phi$ rapidity distributions. The solid points refer to measured 
data, the open points are reflected at midrapidity. The full lines show
the parametrization by a single Gaussian (\ref{eq:rapgauss}), the
dashed lines that by the sum of two Gaussians (\ref{eq:rap2gauss}).
The data at 158$A$~GeV are compared with previously published results
of NA49~\cite{na492000} and CERES~\cite{adamova2006}.
Only statistical errors are shown.}
\label{fig:yspectra}
\end{figure*}

Figure~\ref{fig:yspectra} shows the rapidity distributions, which for
all five energies are in good agreement with both the single-Gaussian
and the double-Gaussian parametrization (see curves). Numerical data
are given in Table~\ref{tab:data}. For the data
sets at 20$A$ and 30$A$~GeV, due to the low number of data points, 
the double-Gaussian fit was constrained to
$a = \sigma_y$ as suggested by the data at 40$A$ and 80$A$~GeV.

Only at 80$A$~GeV is the complete forward hemisphere covered. 
At 158$A$~GeV, large rapidities are not
measured because of the upper momentum cut on the secondary kaons.
Since kaons below 2 GeV laboratory momentum cannot be reliably identified
by $dE/dx$ because of the crossing of the Bethe-Bloch curves, no signal
could be extracted at midrapidity for the lower three beam energies.
The uncertainties in the extrapolation toward midrapidity is
demonstrated by the difference of the two parametrizations. It adds
to the systematic error of the total yield and, in particular,
to that of $dn/dy$ at midrapidity. 
Table~\ref{tab:yspectra} lists the parameters
obtained by the two fit functions, respectively.

Alternatively, the rapidity distributions can be characterized 
by their second moments in a model-independent fashion. The root mean square
of the distributions was calculated from the measured data and
extrapolated to the full rapidity range using the 
parametrizations~(\ref{eq:rapgauss}) and~(\ref{eq:rap2gauss}). The average
of the two results is listed in Table~\ref{tab:yspectra}.

\begin{table*}
\caption{\label{tab:yspectra}Parameters of the single-Gauss fit (\ref{eq:rapgauss})
and the double-Gauss fit (\ref{eq:rap2gauss}) 
to the rapidity distributions. The RMS was calculated from the
data points and extrapolated to the full rapidity range using the
average of the two parametrizations. The first error is statistical,
the second one systematic.}
\begin{center}
\begin{ruledtabular}
\begin{tabular}{ccccccc}
$p_{\rm beam}$  & $\sigma_1$ & $\chi_1^2 / \rm{ndf}$ &
$\sigma_2$ & $a$ & $\chi_2^2 / \rm{ndf}$ & ${\rm rms}_y$ \\
($A$~GeV) \\
\hline
20  &  $0.572 \pm 0.037 \pm 0.030$  &  0.042/2  &
$0.425 \pm 0.026 \pm 0.022$  & 0.425 &  0.75/2  &
$0.582 \pm 0.031 \pm 0.040$  \\

30  &  $0.752 \pm 0.047 \pm 0.057$  &  2.02/3  &
$0.538 \pm 0.028 \pm 0.032$  &  0.538  &  1.03/3  & 
$0.769 \pm 0.030 \pm 0.062$ \\

40  &  $0.863 \pm 0.033 \pm 0.042$  &  3.14/4  &
$0.696 \pm 0.118 \pm 0.036$  &  $0.487 \pm 0.149 \pm 0.051$  &  3.10/3 &
$0.852 \pm 0.015 \pm 0.038$ \\

80  &  $1.016 \pm 0.028 \pm 0.033$  &  17.55/6  &
$0.658 \pm 0.035 \pm 0.043$  &  $0.682 \pm 0.029 \pm 0.043$  &  4.12/5 &
$0.974 \pm 0.024 \pm 0.074$
 \\
158 &  $1.451 \pm 0.086 \pm 0.012$  &  2.36/7 &
& & &
$1.444 \pm 0.021 \pm 0.054$ \\
\end{tabular}
\end{ruledtabular}
\end{center}
\end{table*} 

Total yields were obtained by summation of the data points in the rapidity
spectra and extrapolation to the full rapidity range by the average of
the fit functions. The midrapidity yield $dn/dy$ was calculated 
analytically from the average of the fit functions. For the determination 
of statistical and systematic errors, the correlation of the spectral 
parameters were properly taken into account. The results for the mean multiplicity
of $\phi$ mesons $\langle \phi \rangle$ and for the midrapidity yield $dn_\phi/dy$ 
are listed in Table~\ref{tab:yields}.

\begin{table}
\caption{\label{tab:yields}Total $\phi$ multiplicity $\langle \phi \rangle$
and midrapidity yield $dn_\phi/dy$
calculated from the rapidity distributions of Fig.~\ref{fig:yspectra}.
The first error is statistical, the second one systematic.}
\begin{ruledtabular}
\begin{tabular}{ccc}
$p_{\rm beam}$ ($A$~GeV)  &  $\langle \phi \rangle$  &  $dn_\phi/dy (y_{\rm c.m.})$ \\
\hline
20  &  $1.89 \pm 0.31 \pm 0.22$  &  $1.17 \pm 0.23 \pm 0.38$ \\
30  &  $1.84 \pm 0.22 \pm 0.29$  &  $0.94 \pm 0.13 \pm 0.30$ \\
40  &  $2.55 \pm 0.17 \pm 0.19$  &  $1.16 \pm 0.16 \pm 0.14$ \\
80  &  $4.04 \pm 0.19 \pm 0.31$  &  $1.52 \pm 0.11 \pm 0.22$ \\
158 &  $8.46 \pm 0.38 \pm 0.33$  &  $2.44 \pm 0.10 \pm 0.08$ \\
\end{tabular}
\end{ruledtabular}
\end{table}

All results obtained at 158$A$~GeV are consistent within statistical errors
with NA49 results published earlier~\cite{na492000}, which were obtained
from a data set taken in 1995 (squared symbols in Fig.~\ref{fig:yspectra}(e). 
The main difference of the
two data sets is an improved $dE/dx$ resolution, resulting in a reduced
pion contamination of the kaon candidate sample. The cleaner kaon
identification reduces the distortions in the background-subtracted
invariant-mass spectrum induced by resonances with a pion as decay
daughter~\cite{friese1999}, thus leading to a smaller systematic error
of the Breit-Wigner fit to the spectrum. We thus prefer to use the newly
obtained results at 158$A$~GeV for the discussion.

\section{Discussion}

\begin{figure}
\includegraphics[width=0.45\linewidth]{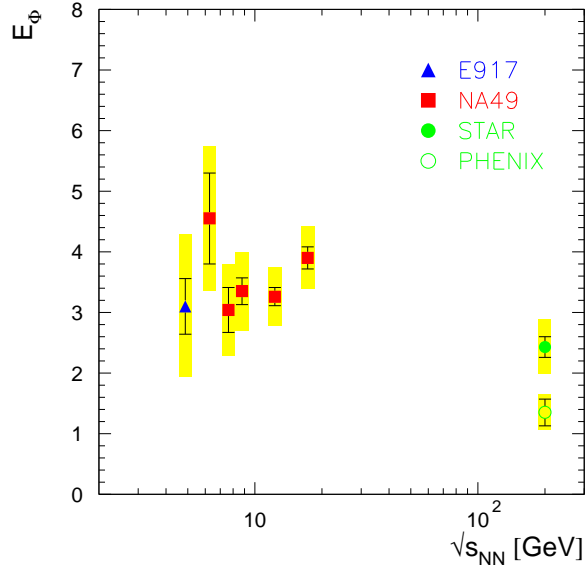}
\caption{(Color online) $\phi$ enhancement factor $E_\phi$ 
[see Eq.~(\ref{eq:enhancement})] as function of energy per nucleon pair. 
Data from the AGS~\cite{back2004} and the SPS refer to
multiplicities in full phase space, data from the RHIC~\cite{adams2005,adler2005}
to midrapidity yields. The shaded boxes represent the systematic errors.}
\label{fig:enhancement}
\end{figure}

The enhancement of relative strangeness production 
in heavy-ion collisions 
with respect to proton-proton reactions is a well-known fact. In an earlier
publication~\cite{na492000}, the enhancement factor for the $\phi$ meson
at top SPS energy was found to be $3.0 \pm 0.7$, thus larger than for 
kaons and $\Lambda$, but smaller than for multistrange hyperons. 
We calculate the $\phi$ enhancement by normalizing the measured $\phi$ meson
yield in $A+A$ by the number of wounded nucleon pairs and dividing
by the corresponding yield in $p+p$. For the lower beam energies, no
reference measurements in elementary collisions are available. Here, we employ
a parametrisation of the $\phi$ excitation function in $p+p$ collisions
as described in Ref.~\cite{back2004}. For top SPS energy and RHIC, the
$\phi$ meson yield measured in $p+p$~\cite{na492000,adams2005} was used. 
Figure~\ref{fig:enhancement} shows the resulting enhancement factor 
\begin{equation}
E_\phi := \frac { 2 \, \langle \phi \rangle_{A+A} } 
{ N_w \, \langle \phi \rangle_{p+p} }
\label{eq:enhancement}
\end{equation} 
as a function of energy per nucleon pair. The measurement
of the E917 Collaboration at AGS ($p_{\rm beam}$ = 11.7$A$~GeV) was extrapolated to full
phase space assuming the same rapidity distribution as for $K^-$ as
suggested by the authors~\cite{back2004}. 
In the AGS/SPS energy region, the value of $E_\phi$ lies between 3 and 4,
and within our experimental uncertainties we find no systematic variation here.
At RHIC energies, the enhancement appears to be lower, significantly so,
should the PHENIX result be validated.
It should be noted, however, that the RHIC values were 
derived from midrapidity data while at lower
energies phase-space integrated yields were used.

In the context of statistical models, the enhancement of strangeness
production can be interpreted as a result of the release of suppression
due to strangeness conservation when going from small ($p+p$)
to large (central $A+A$) systems. Technically, this is reflected in the
application of the canonical ensemble for small systems, while large
systems can be described by the grand-canonical ensemble. In this picture,
a smaller enhancement at RHIC energies points to the fact that at
such high energies, strangeness is produced with sufficient abundance
for the canonical suppression to be relaxed even in $p+p$ collisions.
However, in a purely hadronic picture, canonical suppression does not
act on the $\phi$ meson because it is a strangeness-neutral hadron. 
Enhanced $\phi$ production can thus be attributed either to
enhanced strangeness production in a partonic stage of the collision
or to the coalescence of kaons which suffer canonical suppression
also in a hadronic scenario.

\begin{figure}
\includegraphics[width=0.45\linewidth]{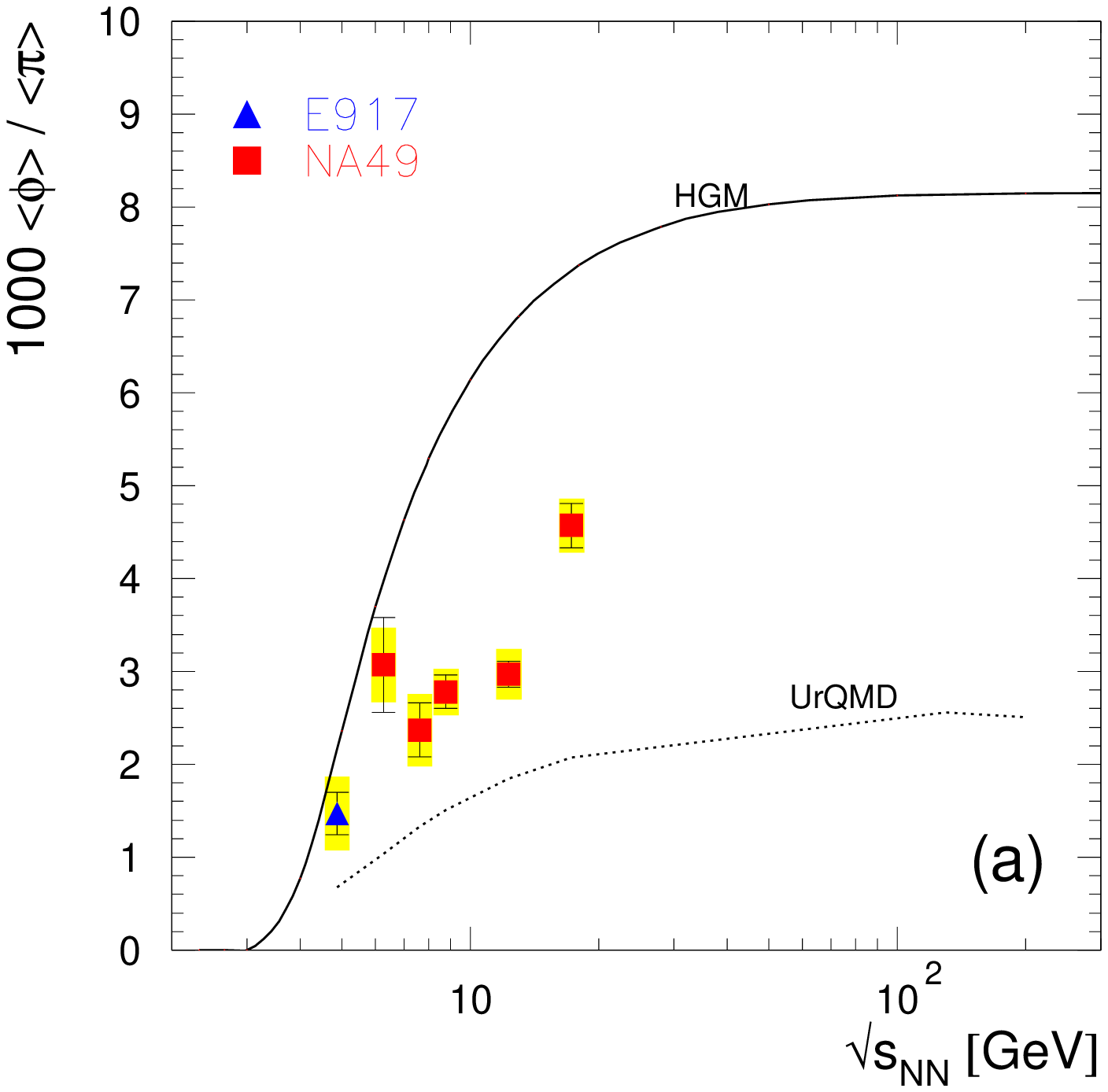}
\includegraphics[width=0.45\linewidth]{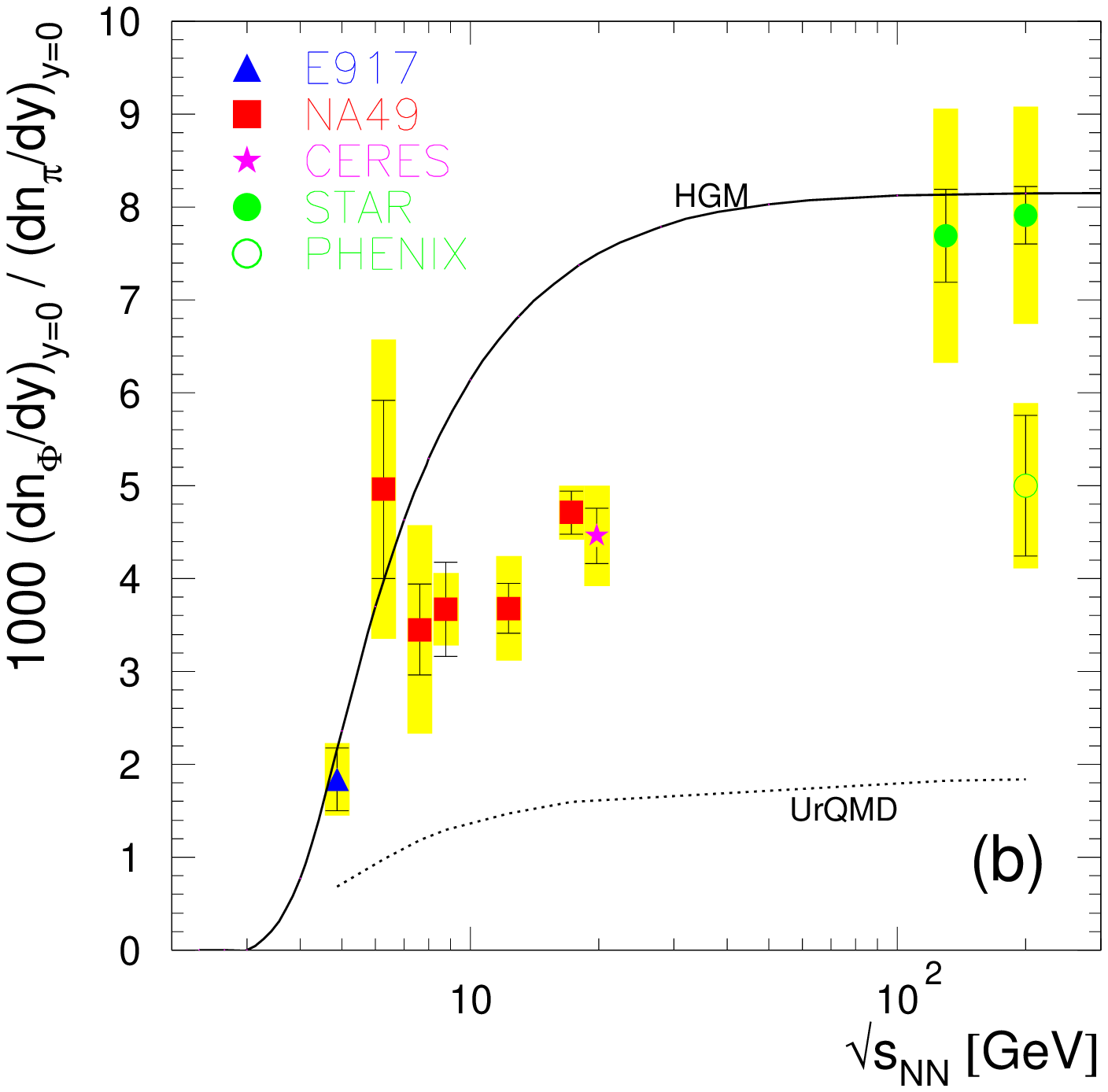}
\caption{(Color online) $\langle \phi \rangle / \langle \pi \rangle$ ratio (a) in full phase
space and (b) at midrapidity as function
of energy per nucleon pair [$\langle \pi \rangle = 1.5 ( \langle \pi^+ \rangle
+ \langle \pi^- \rangle ) $].  The CERES data point~\cite{adamova2006}
was displaced horizontally for visibility. Note that the CERES measurement
is at $y \approx y_{\rm c.m.} - 0.5$.
The full line
shows the predictions of the extended hadron gas model (HGM) with strangeness
equilibration~\cite{andronic2006}, the dashed curves those obtained
with UrQMD 1.3~\cite{urqmd}. The shaded boxes represent the systematic errors.}
\label{fig:phi2pi}
\end{figure}

The hadrochemical models have been extended not only to fit hadron
multiplicities for a given reaction but also to describe the energy
dependence of particle yield ratios by a smooth variation of the
relevant parameters $T$ and $\mu_B$ with collision 
energy~\cite{pbm2002,andronic2006}. Here, the energy dependence of 
temperature and baryochemical potential is obtained by a parametrization
of the values for $T$ and $\mu_B$ 
obtained from fits to particle yield ratios at
various collision energies. The model reproduces 
many yield ratios of the bulk hadrons; however, this does not
hold for the $\phi$ meson, as shown in Fig.~\ref{fig:phi2pi}(a), where
the measured excitation function of the
$\langle \phi \rangle / \langle \pi \rangle$ ratio [$\langle \pi \rangle
= 1.5 (\langle \pi^+ \rangle + \langle \pi^- \rangle)$] is compared
with the model prediction. The relative $\phi$ meson yields at the SPS are 
overpredicted by factors of up to 2. The situation remains essentially unchanged
when midrapidity ratios are considered instead of integrated
yields [Fig.~\ref{fig:phi2pi}(b)].
At the RHIC, there is a large experimental ambiguity as a result of the different
results on $\phi$ production obtained by the STAR and PHENIX
experiments~\cite{adams2005,adler2005}.

\begin{figure}
\includegraphics[width=0.45\textwidth]{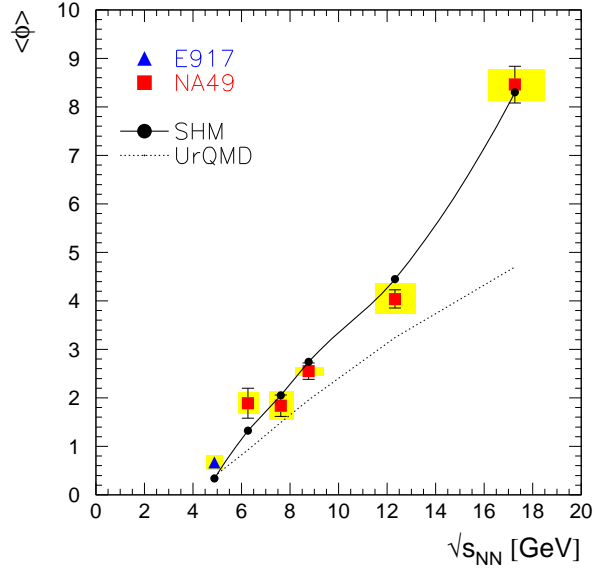}
\caption{(Color online) $\phi$ multiplicity in central $A+A$ collisions as function
of energy per nucleon pair.
The solid points denote
the results of the statistical hadronization model (SHM) which 
allows a deviation from strangeness equilibrium~\cite{becattini2006}.
They are connected by the solid line to guide the eye.
The dotted curve shows the $\phi$ yield predicted by the 
UrQMD~1.3 model~\cite{urqmd}. The shaded boxes represent the systematic errors.}
\label{fig:phiyield}
\end{figure}

A better description of the data is obtained if a strangeness saturation
parameter $\gamma_s$ is allowed. The corresponding model 
predictions~\cite{becattini2006} for the $\phi$ multiplicity, 
resulting from a fit to the hadron abundances at 11.7$A$, 30$A$, 40$A$, 
80$A$, and 158$A$~GeV, are compared with the data in Fig.~\ref{fig:phiyield}
(solid points). Note that this model does not provide a continuous
description of the energy dependence; the points are only connected to guide
the eye.
The agreement with the measurements at the higher 
SPS energies is very good.  The successful 
application of the saturation parameter $\gamma_s$
on the strangeness-neutral $\phi$ meson for $p_{\rm beam} \ge 40A$~GeV 
again suggests that the strangeness content at chemical freeze-out is 
determined on a partonic level for these energies.

Final state interactions after chemical freeze-out could change the 
equilibrium $\phi$ yield and spectra. In particular, scattering of the daughter
kaons with other produced hadrons would lead to a loss of the $\phi$
signal in the experimentally observed decay channel, predominantly
at small rapidities and low values of $p_t$.
Such a loss is not expected in the leptonic decay modes, since electrons
or muons will leave the fireball without interaction.
A comparison of the
measured $m_t$ spectrum via the $K^+K^-$ and $e^+e^-$ decay channels
[see Fig.~\ref{fig:mtspectra}(a)] indicates that the effect cannot be large.
To study the effect on the total yield, we used the string-hadronic transport model
UrQMD~\cite{urqmd}. It was found that only about 8\% of the decayed
$\phi$ mesons are lost for detection due to rescattering of their
daughter particles, independent of collision energy. Similar
results have been obtained with the RQMD model~\cite{johnson2001}. 
The effect is thus not sufficient to account for the deviation of the 
relative $\phi$ multiplicities from their equilibrium values.

On the other hand, $\phi$ mesons can be produced by $KK$ scattering.
In fact, kaon coalescence is the dominant ($\approx$ 70\%) 
production mechanism for the $\phi$ in UrQMD, again for all 
investigated collision systems. As shown by the dotted curve 
in Fig.~\ref{fig:phiyield}, the model gives a reasonable description
of the $\phi$ meson yields at lower energies, whereas it starts to deviate
from the measurements at intermediate SPS energies. The discrepancy
with data is more pronounced when studying the 
$\langle \phi \rangle / \langle \pi \rangle$ ratio 
(Fig.~\ref{fig:phi2pi}) because UrQMD overestimates
the pion yields at SPS energies by about 30\%.

\begin{figure}
\includegraphics[width=0.45\linewidth]{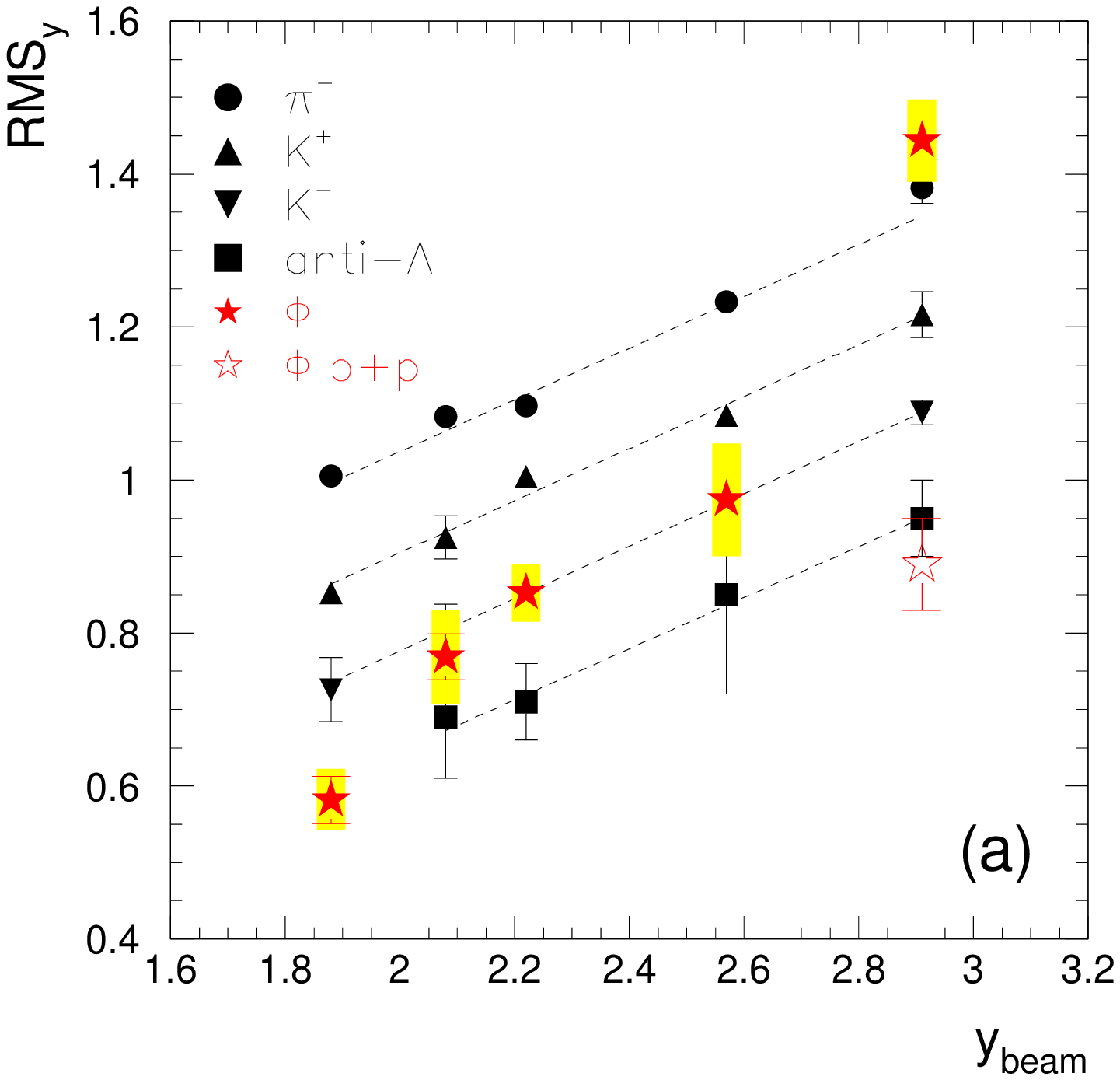}
\includegraphics[width=0.45\linewidth]{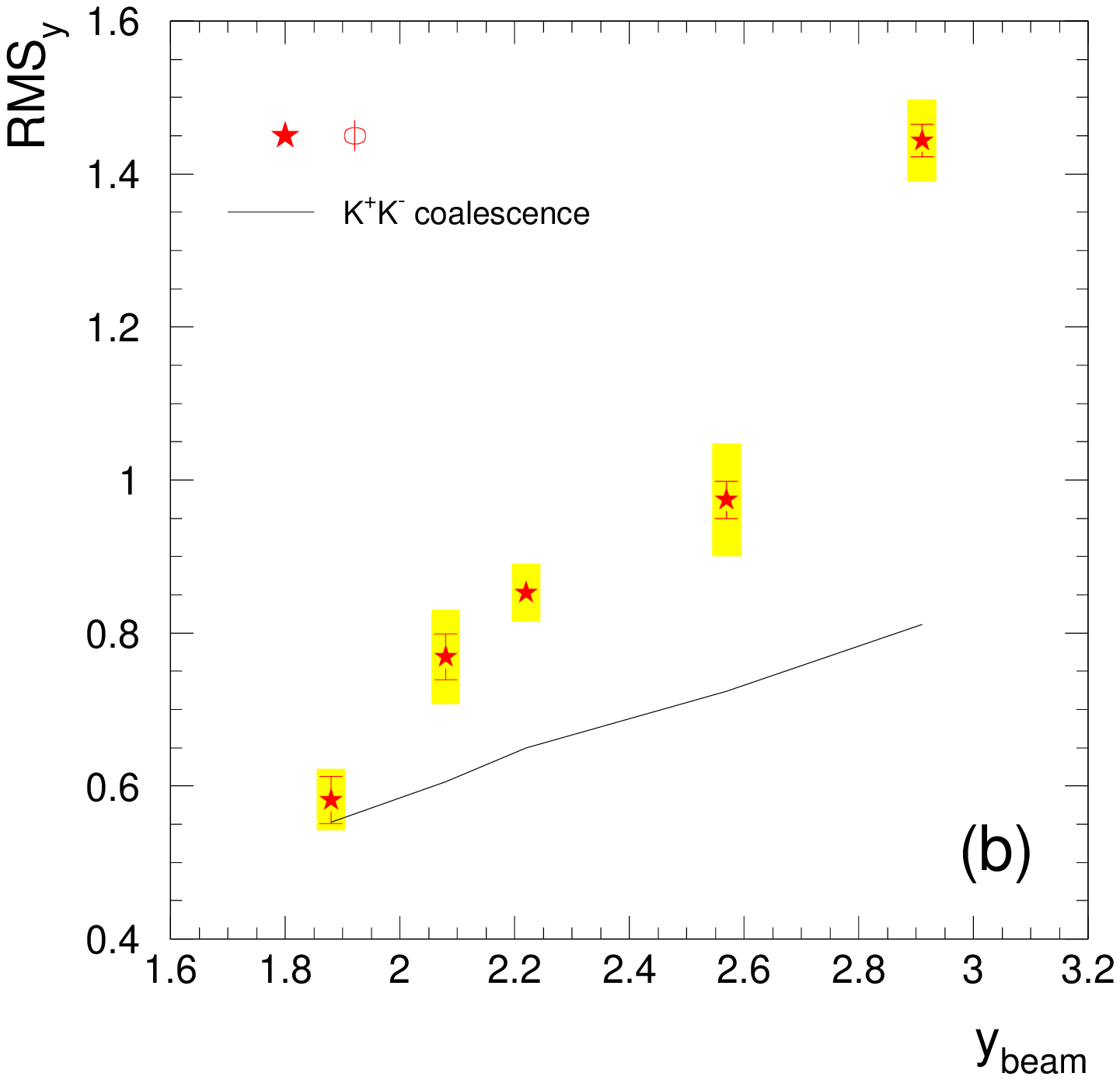}
\caption{(Color online) (a) Widths of the rapidity distributions of $\pi^-$, $K^+$, $K^-$,
and $\phi$ in central Pb + Pb collisions
at SPS energies as function of beam rapidity
\cite{na492002,na492007}. 
The dashed lines are to guide the eye. 
The open star denotes the $\phi$ rapidity width measured in $p+p$ 
collisions~\cite{na492000}.
(b) Widths of the $\phi$ rapidity distributions in central Pb+Pb 
collisions compared with the
expectations in a kaon coalescence picture [Eq.~(\ref{eq:coalescence})].
The shaded boxes represent the systematic errors (shown only for $\phi$ mesons).}
\label{fig:rapwidth}
\end{figure}

The hypothesis that the $\phi$ meson is produced predominantly
by kaon coalescence can be tested by comparing the $\phi$ and
kaon distributions in phase space. Figure~\ref{fig:rapwidth}(a)
shows the width of the $\phi$ rapidity distribution as a function
of beam rapidity at SPS energies, together with that measured
for $\pi^-$, $K^+$, and $K^-$ \cite{na492002,na492007}.
The $\phi$ meson width does not fit into the systematics
observed for the other particle species but
increases much faster with energy.
While at 20$A$~GeV, the $\phi$ rapidity distribution is narrower 
than that of $K^-$, we find it at top SPS energy
comparable to the pions. In addition, at 158$A$~GeV it is much larger
in central Pb + Pb collisions than measured 
in $p+p$ collisions
at the same energy~\cite{na492000}, a feature which is not observed
for other particle species.

In the kaon coalescence picture, there would be a tendency for the
$\phi$ rapidity distribution to be narrower than those of the kaons.
In an ideal case, neglecting correlations, 
\begin{equation}
\frac {1} {\sigma^2_\phi} = 
\frac {1} {\sigma^2_{K^+}} + \frac {1} {\sigma^2_{K^-}}  \, ,
\label{eq:coalescence}
\end{equation}
where the distributions were approximated by Gaussians. As shown
in Fig.~\ref{fig:rapwidth}(b), the $\phi$ data
rule out kaon coalescence 
as dominant formation mechanism for beam energies above 30$A$~GeV.
Only at 20$A$~GeV, the observed rapidity widths are consistent
with the coalescence picture.
As mentioned before, this would also explain the $\phi$ enhancement
at low energies, where a transient deconfined state is not expected.

The observation that models based on a purely hadronic reaction scenario
have serious problems in describing relative strangeness production
in the upper SPS energy range is not unique to the $\phi$ meson
but holds for kaons and other strange particles, too. It has been related
to the onset of deconfinement at around 30$A$~GeV as predicted
by the statistical model of the early stage~\cite{gazdzicki1999}.
A striking experimental evidence is the narrow maximum in the 
$K^+ / \pi^+$ ratio at this energy~\cite{na492002,na492007}.
A similar structure is, within experimental errors,
not observed for the $\phi$ meson (Fig.~\ref{fig:phi2pi});
instead, the energy dependence of the relative $\phi$ meson yield
resembles that of the $K^-$. 
This can be understood since the $K^+$ yield is in good approximation 
proportional to the total strangeness production, which is not the
case for $K^-$ and $\phi$ because a large, energy-dependent fraction
of $s$ quarks is carried by hyperons.

\begin{figure}
\includegraphics[width=0.45\linewidth]{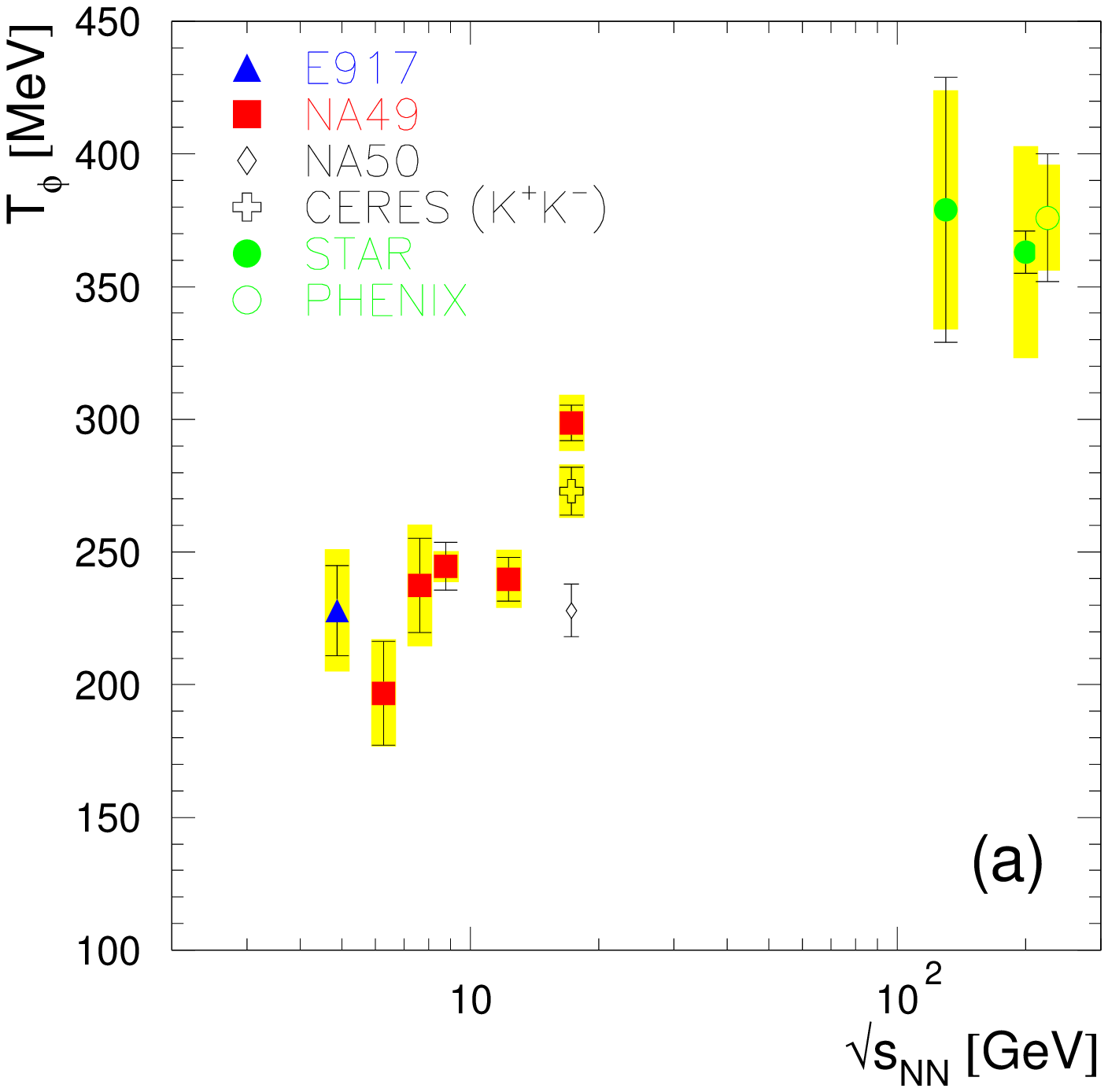}
\includegraphics[width=0.45\linewidth]{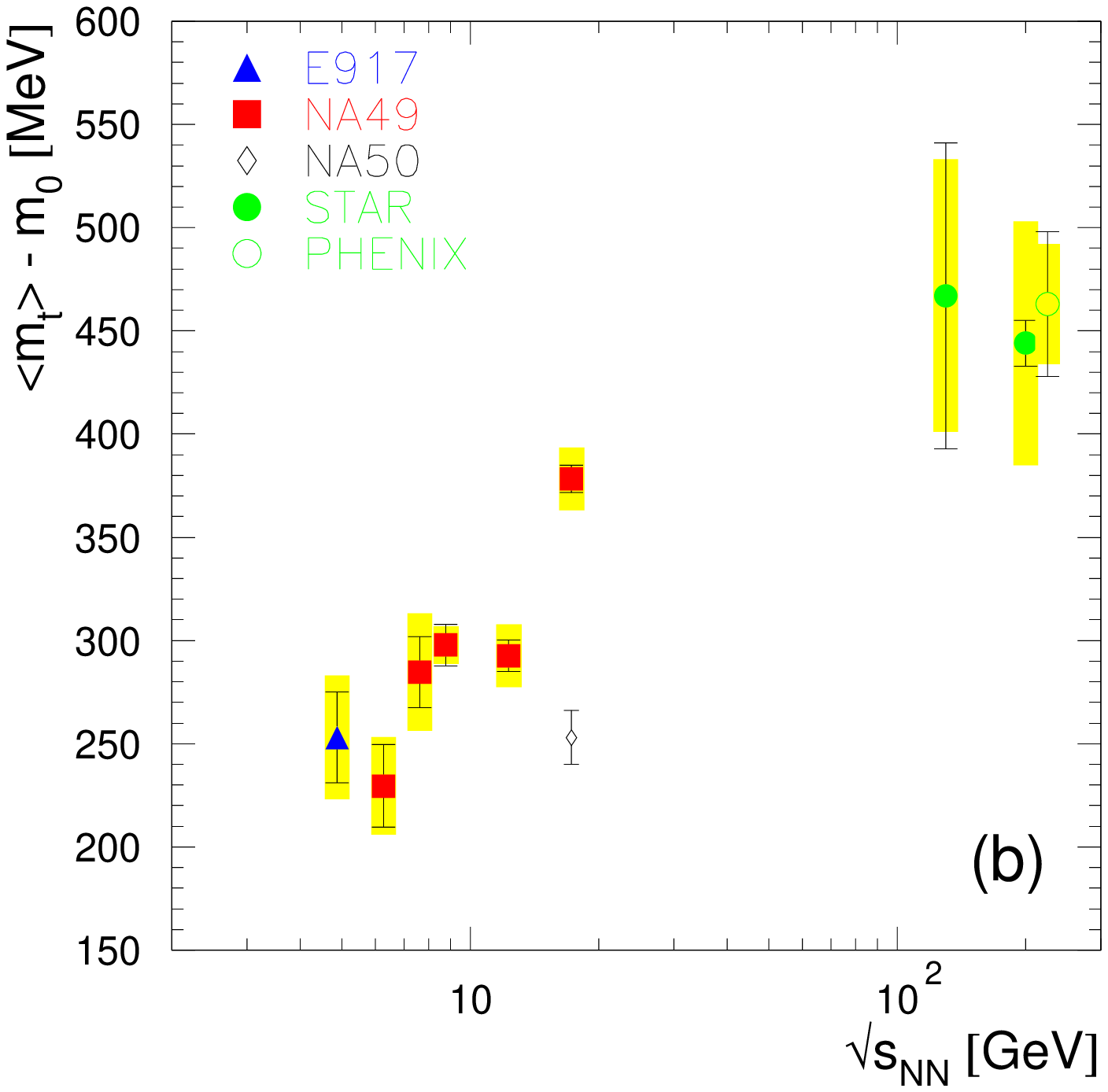}
\caption{(Color online) (a) Inverse slope parameter $T$ and (b) average transverse mass
$\langle m_t \rangle - m_0$ of the $\phi$ meson in central $A+A$
collisions as function of energy per nucleon pair. The data from E917~\cite{back2004} were averaged over the measured
rapidity interval (see Table~\ref{tab:ptspectra}). Results from NA50~\cite{alessandro2003} and
RHIC~\cite{adler2003,adams2005,adler2005} were obtained at
midrapidity, the result from CERES at $y = - 0.71$. Data from NA49 are integrated over rapidity. The PHENIX
data point was slightly displaced horizontally for visibility. 
For the NA49 data, $\langle m_t \rangle$ was calculated from the 
transverse momentum spectra using an exponential extrapolation to full
$p_t$. For the other data sets, it was derived analytically from
the exponential fit function. The shaded boxes represent the systematic errors.}
\label{fig:slopes}
\end{figure}

The energy dependences of both the inverse slope parameter and
the mean transverse mass of the $\phi$ meson are shown in 
Fig.~\ref{fig:slopes}.
The transverse mass spectra of the $\phi$
are well described by exponential fits 
[see Fig.~\ref{fig:mtspectra}(a)]; consequently, the two parameters show
a similar behavior.
Over the energy range AGS--SPS--RHIC, there is an overall tendency for both
parameters to increase. However, a constancy of the values in the lower
SPS energy range, as has been observed for pions, kaons, and 
protons~\cite{na492007}---a fact interpreted as being consistent
with a mixed partonic/hadronic phase~\cite{gorenstein2003}---cannot be excluded.


\section{Summary}

We have presented new data on $\phi$ production in central Pb+Pb
collisions obtained by the NA49 experiment at 20$A$, 30$A$, 40$A$, 80$A$,
and 158$A$~GeV beam energies.
No indications of medium modifications of the $\phi$ meson mass or width
were observed. The energy dependence of the production
characteristics was studied by comparing them with measurements at AGS
and RHIC energies. We find that at low SPS energy, the data can be
understood in a hadronic reaction scenario; while at higher energies,
hadronic models fail to reproduce the data. A statistical hadron 
gas model with undersaturation of strangeness gives a good description 
of the measured yields. This suggests that $\phi$ production is
ruled by partonic degrees of freedom,
consistent with the previously found indications for the onset of 
deconfinement at lower SPS energy.

\section*{Acknowledgments}
This work was supported by the U.S.~Department of Energy
Grant DE-FG03-97ER41020/A000,
the Bundesministerium f{\"u}r Bildung und Forschung, Germany (06F137),
the Polish State Committee for Scientific Research
(2 P03B 006 30, SPB/CERN/P-03/Dz 446/2002-2004, 2 P03B 04123),
the Hungarian Scientific Research Foundation (T032648, T032293, T043514),
the Hungarian National Science Foundation, OTKA, (F034707),
the Polish-German Foundation,
and the Korea Research Foundation Grant (KRF-2003-070-C00015).

\end{document}